\documentclass[%
reprint,
superscriptaddress,	
groupedaddress,
unsortedaddress,
amsmath,amssymb,
prx,
noeprint,
]{revtex4-2}

\usepackage{graphicx}
\graphicspath{{./figs/}{../figs/}}  
\usepackage{bm}
\usepackage{color}
\usepackage{hyperref}
\usepackage{siunitx}

\def\dd{{\mathrm{d}}}
\def\jj{j}
\def\pdv#1#2{\frac{\partial#1}{\partial#2}}
\def\odv#1#2{\frac{\dd#1}{\dd#2}}

\usepackage{amsmath}
\begin{document}


\title{Hidden Symmetry Protection and Topology in Surface Maxwell Waves }
\author{Yosuke Nakata}
\email{y.nakata.es@osaka-u.ac.jp}
\affiliation{Graduate School of Engineering Science, Osaka University, Osaka 560-8531, Japan}
 \affiliation{Center for Quantum Information and Quantum Biology, Osaka University, Osaka, 560-8531, Japan}
\author{Toshihiro Nakanishi}
\affiliation{%
Department of Electronic Science and Engineering, Kyoto University, Kyoto 615-8510, Japan
}
\author{Ryo Takahashi}
\affiliation{%
Advanced Institute for Materials Research (AIMR), Tohoku University, Miyagi 980-8577, Japan
}
\author{Fumiaki Miyamaru}
\affiliation{%
Department of Physics, Faculty of Science, Shinshu University, Nagano 390-8621, Japan
}
\author{Shuichi Murakami}%
\affiliation{%
Department of Physics, Tokyo Institute of Technology, Tokyo 152-8551, Japan
}

\date{\today}

\begin{abstract}
Since the latter half of the 20th century, the use of metal in
optics has become a promising plasmonics field for controlling light at
a deep subwavelength scale. Surface plasmon polaritons localized on 
metal surfaces are crucial in plasmonics. However, despite the long history
of plasmonics, the underlying mechanism producing the surface waves is not fully understood. 
This study unveils the hidden symmetry protection that ensures the existence of degenerated electric zero modes.
These zero modes are identified as physical origins of surface plasmon polaritons,
and similar zero modes can be directly excited at a temporal boundary. 
In real space, the zero modes possess vector-field rotation related to surface impedance.
Focusing on the surface impedance, 
we prove the bulk--edge correspondence, which guarantees the existence of surface plasmon polaritons even with nonuniformity.
Lastly, we extract the underlying physics in the topological transition between metal and dielectric material using a minimal circuit model with duality.
The transition is considered the crossover between electric and magnetic zero modes.
\end{abstract}

\maketitle

\section{\label{sec:introduction} Introduction}

Metal has been one of the fundamental materials for producing optical elements, such as mirrors,
for over 5000 years \cite{Enoch2006}.
However, despite its extensive history, wave propagation inside metals has received little attention, 
because electromagnetic waves are attenuated in metals
owing to their \textit{negative response}.
Because free electrons in a metal are sensitive to oscillating electric fields,
the electric field and induced current have opposite phases. 
Since the 20th century, researchers have investigated extraordinary light propagation enabled by a negative response. 
One prominent example is the discovery of a negative refractive index,
which can be realized in a medium with simultaneous negative responses for electric and magnetic fields \cite{Veselago1968}. 
Remarkably, a negative refractive index can be applied to realize a flat lens, which can help overcome the diffraction limit \cite{Pendry2000}.
Because there is no natural material with a negative refractive index, 
the discovery stimulated the development of artificial materials called \textit{metamaterials} \cite{Caloz2006, Solymar2009},
and negative refraction was eventually demonstrated in a metamaterial \cite{Shelby2001a}.
These findings demonstrate the potential capability of the negative response in optics.

The negative response impacts not only spatial wave propagation but also surface-wave formation.
In fact, a metallic surface supports surface plasmon polaritons, i.e.,
hybridized waves comprising a plasmonic electron oscillation and an electromagnetic wave \cite{Sarid2010, Sarkar2017}.
As surface plasmon polaritons can be squeezed into a deep subwavelength volume, 
they play essential roles in nanophotonics toward the miniaturization of optics,
and the research field involving surface plasmon polaritons is called plasmonics \cite{Maier2007}.
Although surface plasmon polaritons have been studied for over half a century,
the investigation of their origin only began recently.
In the paradigm of topological physics, 
integers are used to characterize bulk materials,
where the bulk--edge correspondence predicts
the existence of a boundary mode between two materials with different topological numbers \cite{asboth2016short,Vanderbilt2018}.
The bulk--edge correspondence provides a powerful guiding principle; however, it is often empirical and requires exact proof for each case.
For plasmonic systems, different integers are used
to distinguish between a metal and dielectric material,
such as the $\mathbb{Z}_4$ number describing the winding of the complex helicity spectrum \cite{Bliokh2019} and the Zak phase \cite{Yang2020}.
The existence of surface plasmon polaritons is indicated by the bulk--edge correspondence.
However, these seminal works are limited to simply observing surface-wave formations and lack rigorous proof of the bulk--edge correspondence.

In this study, special zero modes were identified as the origin of surface Maxwell waves,
and a general bulk--edge correspondence is rigorously established
that explained the existence of surface plasmon polaritons on metals.
The first approach is based on \textit{symmetry protection}.
It guarantees the existence of localized states under certain symmetry.
In Sec.~\ref{sec:sym_protect}, we formulate symmetry-protected \textit{electric} zero modes in electrostatics,
which are the origin of surface plasmon polaritons.
The robustness of symmetry protection in some nonuniform systems is confirmed.
Additionally, we demonstrate that analogous zero modes can be experimentally excited at a temporal boundary.
In Sec.~\ref{sec:topo_rot}, we investigate real-space topological polarization rotation 
in the electric zero modes. 
Keller--Dykhne self-duality is identified as the physical origin of polarization rotation. 
Furthermore, we reveal the relationship between the polarization rotation and surface impedance.
In Sec.~\ref{sec:bulk-edge}, we define topological integers based on the surface impedance 
and establish the bulk--edge correspondence, which provides another way to understand surface plasmon polaritons even with nonuniformity.
In Sec.~\ref{sec:dielectric_metal}, we propose a minimal circuit model to explain the
underlying physics of the continuous transition between metal and dielectric.
From a physical standpoint, the transition is due to the crossover between electric and magnetic zero modes.

\section{\label{sec:sym_protect} Symmetry Protection}

Certain symmetry often guarantees the existence of 
topological end or boundary states. 
For instance, in the Su--Schrieffer--Heeger (SSH) model \cite{Su1979,Su1980},
the sublattice symmetry works as a chiral symmetry and 
protects end states at zero energy \cite{asboth2016short}.
Nevertheless, the generalization of the SSH model in a continuous system 
simply leads to the Dirac electron \cite{Schindler2020}.
The topological boundary state of the Dirac electron appears at the boundary 
between the positive and negative mass regions in the Jackiw--Rebbi model, and its energy is maintained to be zero \cite{Jackiw1976}.

Herein, we construct symmetry-protected electric zero modes in electrostatics
and identify them as the robust origin of surface plasmon polaritons.
Additionally, we show that the analogous zero modes can be excited at a temporal boundary.

\subsection{Mechanism}

We construct symmetry-protected electric zero modes in electrostatics
and identify their boundary degree of freedom.
Finally, we discuss the origin of the singular charge response 
near a surface with symmetry.

\paragraph{Setup. }
Consider an electrostatic problem described by a scalar permittivity $\varepsilon(x,y,z)$.
Without free charge, the fundamental equations are given as follows:
\begin{equation}
 \nabla \cdot \mathbf{D} =0,\quad \nabla\times \mathbf{E}=0, \label{eq:1}
\end{equation}
where $\mathbf{D}$ and $\mathbf{E}$ represent the electric displacement and electric field, respectively.
The constitutive relation is expressed as follows:
\begin{equation}
 \mathbf{D}(x,y,z) = \varepsilon(x, y, z) \mathbf{E}(x,y,z).  \label{eq:2}
\end{equation}
We mainly focus on a particular distribution of $\varepsilon(x, y, z)$ that satisfies the following equation:
\begin{equation}
 \varepsilon(-x, y, z) =  -\varepsilon(x, y, z).  \label{eq:3}
\end{equation}
For simplicity, we assume that $\varepsilon(x, y, z) > 0 $ in $x\geq 0^+$
and that there is no free charge unless noted otherwise.
To handle discontinuous functions, such as $\varepsilon$, 
we sometimes distinguish the positive side of zero ($0^+=\lim_{x\rightarrow 0,\ x>0} x$) from the negative side ($0^-=\lim_{x\rightarrow 0,\ x<0} x$).

\paragraph{Symmetry Operations.}

We introduce two symmetry operations to characterize Eq.~(\ref{eq:3}).
First, we consider the mirror reflection $\mathcal{M}_x$ with respect to the plane $x=0$.
Under the $\mathcal{M}_x$ operation, a polar-vector field $\mathbf{F}(x,y,z)=[F_x(x,y,z)\ F_y(x,y,z)\ F_z(x,y,z)]^\mathrm{T}$ transforms into
$\mathbf{F}'(x,y,z) =  [-F_x(-x,y,z)\ F_y(-x,y,z)\ F_z(-x,y,z)]^\mathrm{T}$.
This transformation can also be expressed as $\mathbf{F}' = \mathcal{M}_x \mathbf{F}$.
To preserve Eqs.~(\ref{eq:1}) and (\ref{eq:2}) under $\mathcal{M}_x$, the permittivity should be expressed as follows when considering transformed fields $\mathbf{E}'=\mathcal{M}_x \mathbf{E}$ and $\mathbf{D}'=\mathcal{M}_x \mathbf{D}$:
\begin{equation}
 \varepsilon'(x,y,z) = (\mathcal{M}_x \varepsilon)(x,y,z) = \varepsilon(-x,y,z)  \label{eq:4}
\end{equation} 

The second operation is on the internal degree of freedom between $\mathbf{E}$ and $\mathbf{D}$.
Consider the following conjugation operation $\mathcal{C}$ for $(\mathbf{E},\mathbf{D})$:
\begin{equation}
\mathcal{C} (\mathbf{E},\mathbf{D}) = (\mathbf{E}, -\mathbf{D}).  \label{eq:5}
\end{equation}
The transformed $(\mathbf{E}', \mathbf{D}')=\mathcal{C} (\mathbf{E},\mathbf{D})$ satisfies Eq.~(\ref{eq:1}).
To preserve Eq.~(\ref{eq:2}) under the $\mathcal{C}$ operation, the permittivity changes as follows:
\begin{equation}
 \varepsilon'(x,y,z) = (\mathcal{C} \varepsilon) (x,y,z) = - \varepsilon(x,y,z).  \label{eq:6}
\end{equation}

The combined operation $\mathcal{CM}_x$ induces permittivity transformation $(\mathcal{CM}_x)\varepsilon(x,y,z) = - \varepsilon(-x,y,z)$.
Thus, Eq.~(\ref{eq:3}) represents the $\mathcal{CM}_x$ symmetry.
Evidently, $(\mathcal{CM}_x)^2$ is identical to the operation $\mathrm{Id}$.
Therefore, the solutions of a system with $\mathcal{CM}_x$ symmetry are classified as follows:
symmetric ($S$) and antisymmetric ($A$) fields:
\begin{align}
 \mathcal{CM}_x (\mathbf{E}_S, \mathbf{D}_S)&= (\mathbf{E}_S, \mathbf{D}_S),  \label{eq:7}\\
 \mathcal{CM}_x (\mathbf{E}_A, \mathbf{D}_A)&= -(\mathbf{E}_A, \mathbf{D}_A).  \label{eq:8}
\end{align}
Here, we regard $-(\mathbf{E},\mathbf{D}):= (-\mathbf{E}, -\mathbf{D})$.

\paragraph{Antisymmetric Solution.}

Furthermore, we show that there is no $\mathcal{CM}_x$-antisymmetric field.
Owing to the antisymmetry and tangential continuity condition, $E_y = E_z =0$ on $x=0$ must follow.
By contrast, antisymmetry yields $E_x(0^-, y, z) = E_x(0^+, y, z)$ and $D_x(0^-, y, z) = -D_x(0^+, y, z)$. Owing to the assumption of no free charge,
$D_x(0^-, y, z) = D_x(0^+, y, z) = 0$ holds true. Therefore, $\mathbf{E}$ and $\mathbf{D}$ on $x=0$ must be zero.
Additionally, all fields must vanish. We can physically justify this statement as follows:
The solution in $x\geq 0^+$ can be safely connected to a vacuum in $x\leq 0^-$, which has $\mathbf{E}=0$ and $\mathbf{D}=0$. Because we have assumed that there is no source in $x\geq 0^+$ with $\varepsilon>0$, we can conclude that all fields in the entire space vanish.

\paragraph{Symmetric Solution.}
A $\mathcal{CM}_x$-symmetric field $(\mathbf{E}, \mathbf{D})$ has a unique feature that \textit{always} satisfies the boundary condition on $x=0$.
This is the most fundamental characteristic of a $\mathcal{CM}_x$-symmetric system.
Note that $(\mathbf{E}, \mathbf{D})$ can be any continuous field and does not need to satisfy Maxwell's equations.
Let us check this special property.
Because $\mathbf{E}$ is symmetric under $\mathcal{M}_x$,
$E_{y}$ and $E_{z}$ are continuous on $x=0$.
Conversely, the electric displacement is antisymmetric under $\mathcal{M}_x$.
Therefore, $D_{x}$ is continuous on $x=0$.
Thus, both of the boundary conditions are automatically satisfied.

\paragraph{$\mathcal{CM}_x$ Symmetrization.}
The above remarkable continuity of a $\mathcal{CM}_x$-symmetric field can be used to obtain a whole-space solution from a half-space solution.
If we have a solution $(\mathbf{E}, \mathbf{D})$ satisfying Eqs.~(\ref{eq:1}) and (\ref{eq:2}) of an electrostatic problem only in $x\geq 0^+$,
the field in $x\leq 0^-$ is constructed via $\mathcal{CM}_x$ symmetrization:
\begin{equation}
(\mathbf{E}, \mathbf{D})(x,y,z) = \big(\mathcal{CM}_x(\mathbf{E}, \mathbf{D})\big)(-x,y,z) \quad (x\leq 0^-). \label{eq:9}
\end{equation}
Here, we abbreviate $(\mathbf{E}, \mathbf{D})(x,y,z)=(\mathbf{E}(x,y,z), \mathbf{D}(x,y,z))$.
From the above field continuity, the boundary condition is automatically satisfied.

\paragraph{$\mathcal{CM}_x$ Point and Dipole Fields.}

\begin{figure}[!t]
 \centering
  \includegraphics{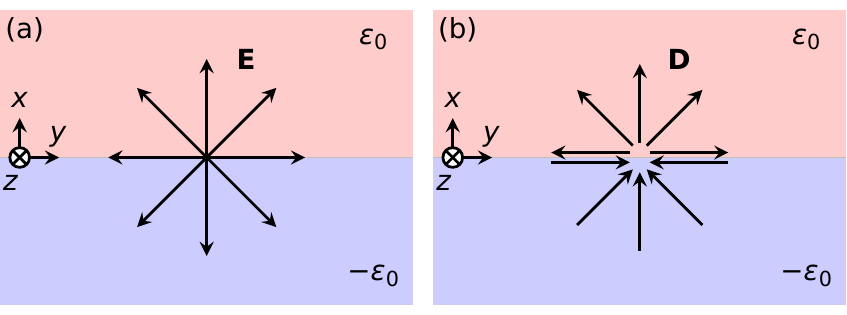}
  \caption {$\mathcal{CM}_x$ point fields: (a)~Electric field. (b)~Electric displacement. 
\label{fig:CMx_single_layer}}
 \end{figure}

\begin{figure}[!t]
 \centering
  \includegraphics{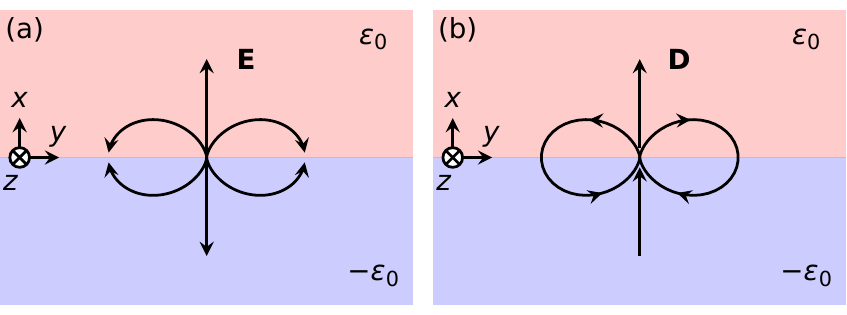}
  \caption{$\mathcal{CM}_x$ dipole fields: (a)~Electric field. (b)~Electric displacement. 
\label{fig:CMx_double_layer}}
 \end{figure}

We introduce the fundamental fields with $\mathcal{CM}_x$ symmetry.
Consider a half-space $\varepsilon(x,y,z)>0$ in $x\geq 0^+$.
We begin with the $\mathcal{M}_x$-symmetrized permittivity, which is defined as follows:
\begin{equation}
 \varepsilon_\mathcal{M}(x,y,z) 
=\begin{cases}
\varepsilon(x,y,z) & (x\geq 0)\\
\varepsilon(-x,y,z) & (x\leq 0)
\end{cases} \label{eq:10}
\end{equation} 
First, we place a point charge $q$ at $(x,y,z)=(0^-,Y,Z)$ in 
$\varepsilon_\mathcal{M}$.
The $\mathcal{CM}_x$ symmetrization is applied to the field in $x\geq 0^+$ to eliminate the point source.
Under symmetrization, the permittivity becomes $\mathcal{CM}_x$-symmetric. 
The obtained field is called a $\mathcal{CM}_x$ point field and is denoted as 
$(\mathbf{E}_{\mathbf{R}}^\mathrm{(pt)}, \mathbf{D}_{\mathbf{R}}^\mathrm{(pt)})$ 
with $\mathbf{R} = [0\ Y\ Z]^\mathrm{T}$.
The most straightforward situation with a uniform $\varepsilon(x,y,z) = \varepsilon_0$ is 
shown in Fig.~\ref{fig:CMx_single_layer}.
The second example is the dipole field.
Consider a dipole with the dipole moment $(p,0,0)$ at $(x,y,z)=(0^-,Y,Z)$ in $\varepsilon_\mathcal{M}$.
The $\mathcal{CM}_x$ symmetrization for the field in $x\geq 0^+$ removes the dipole source.
The obtained field is called a $\mathcal{CM}_x$ dipole field and is
denoted as $(\mathbf{E}_{\mathbf{R}}^\mathrm{(dp)}, \mathbf{D}_{\mathbf{R}}^\mathrm{(dp)})$ with $\mathbf{R} = [0\ Y\ Z]^\mathrm{T}$.
The $\mathcal{CM}_x$ dipole fields for a uniform $\varepsilon(x,y,z)=\varepsilon_0$ 
are shown in Fig.~\ref{fig:CMx_double_layer}.

\paragraph{Boundary Degree of Freedom.}

We show that the degree of freedom of $\mathcal{CM}_x$-symmetric fields
is represented by either a $\mathcal{CM}_x$ point field or a $\mathcal{CM}_x$ dipole field.
Consider a half-space electrostatic potential $\varphi$ in $x\geq 0^+$ 
with $\varepsilon(x,y,z)>0$.
Assuming that there is no free charge in $x> 0$, 
the boundary charge or dipole may appear at $x=0$.
Let $\varphi_S$ and $\varphi_A$ be 
symmetric and antisymmetric extensions of $\varphi$ in the whole space, respectively.
We define $\varphi_S$ and $\varphi_A$ as follows:
\begin{equation}
 \varphi_S(x,y,z) = 
\begin{cases}
\varphi(x,y,z) & (x\geq 0^+)\\
\varphi(-x,y,z) & (x\leq 0^-)
\end{cases}  \label{eq:11}
\end{equation}
\begin{equation}
 \varphi_A(x,y,z) = 
\begin{cases}
\varphi(x,y,z) & (x \geq 0^+)\\
-\varphi(-x,y,z) & (x \leq 0^-)
\end{cases}   \label{eq:12}
\end{equation}
These potentials satisfy $\nabla \cdot \varepsilon_\mathcal{M} \nabla \varphi =0 $ in $x\ne 0$ with 
the $\mathcal{M}_x$-symmetrized permittivity defined in Eq.~(\ref{eq:10}).
To ensure the boundary condition on $x=0$ for $\varphi_S$,
there should be a boundary charge $\sigma$ on $x=0$ satisfying the following equation: 
\begin{equation}
 \sigma(y,z) = -2\varepsilon(0,y,z) \pdv{\varphi}{x}(0^+,y,z).  \label{eq:13}
\end{equation}
On $x=0$, the tangential component $\mathbf{E}_t$ is continuous, whereas the normal component $D_x$ is discontinuous indicated by $\sigma$.
In fact, $D_x(0^+, y, z) = \sigma(y,z)/2$ holds true.
A nonuniform $\sigma$ may also contribute to the tangential component.
Conversely, $\varphi_A$ has discontinuity on $x=0$, 
which indicates the existence of a double layer. This is described as follows:
\begin{equation}
 \tau_x(y,z) = 2\varepsilon(0,y,z)\varphi(0^+,y,z).  \label{eq:14}
\end{equation}
For the double layer on $x=0$, the normal component $D_x$ is continuous, 
whereas the tangential component $\mathbf{E}_t$ exhibits discontinuity by $\mathbf{E}_t(0^+,y,z) - \mathbf{E}_t(0^-,y,z)=-\nabla [\tau_x(y,z)/\varepsilon(0,y,z)]$ \cite{Stratton1941}. Thus, we obtain $\mathbf{E}_t(0^+, y,z) = -(1/2)\nabla [\tau_x(y,z)/\varepsilon(0,y,z)]$, which agrees with Eq.~(\ref{eq:14}).
Additionally, a nonuniform $\tau_x$ may contribute to the normal component via electric-field leakage to outside the double layer.
From the observation of characteristics of the single and double layers thus far, 
we can conclude that either $\sigma$ or $\tau_x$ can be used to construct the field in $x\geq 0^+$.
This statement is consistent with the treatment 
of a conventional boundary-value problem: the solution to an electrostatic problem is uniquely 
determined by applying either the Dirichlet or Neumann boundary conditions
for each boundary \cite{Jackson1998}.
Now, consider $\Phi = (\varphi_S + \varphi_A)/2$.
This field is expressed as follows:
$\Phi = \varphi$ in $x\geq 0^+$ and $\Phi = 0$ in $x\leq 0^-$; i.e.,
the field in $x\geq 0^+$ is generated 
from $\sigma/2$ and $\tau_x/2$ on $x=0$
in a $\mathcal{M}_x$-symmetrized system.
By contrast, $\Phi$ vanishes in $x\leq 0^-$.
To obtain a $\mathcal{CM}_x$-symmetric solution, we apply 
the $\mathcal{CM}_x$ symmetrization for $\Phi$, which makes both $D_x$ and $\mathbf{E}_t$ continuous on $x=0$.
All sources then vanish, whereas the field remains.

\paragraph{Singular Response.}

We characterize the singular response of a $\mathcal{CM}_x$ system.
Consider a system with permittivities of $\varepsilon_1(x,y,z)$ in $x\geq 0^+$ and $\varepsilon_2(x,y,z)$ in $x\leq 0^-$. 
These permittivities do not need to have $\mathcal{CM}_x$ symmetry.
To construct two modes similar to symmetric and antisymmetric modes,
we assume the following condition:
\begin{equation}
 \frac{\varepsilon_2(-x,y,z)}{\varepsilon_1(x,y,z)} = \mathrm{Const.}\quad (x\geq 0^+),  \label{eq:15}
\end{equation}
which is always satisfactory for a uniform $\varepsilon_1$ and $\varepsilon_2$.
Now, we place charges $q_1$ and $q_2$ at $(x,y,z)=(a,0,0)$ and $(x,y,z)=(-a,0,0)$, respectively, with $a>0$.

\begin{figure}[!t]
 \centering
  \includegraphics{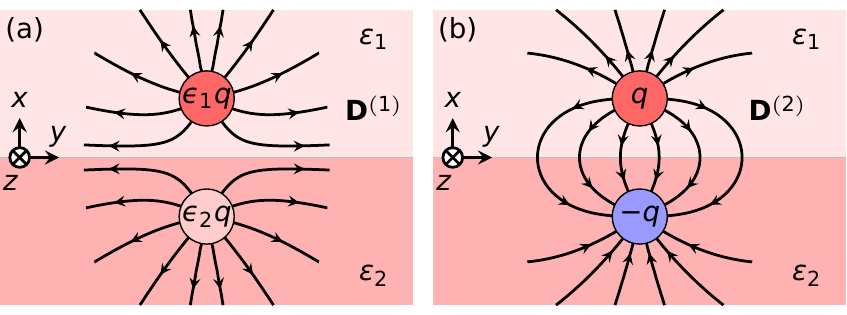}
  \caption{
\label{fig:charge_response} Electric-displacement fields with symmetries:
(a) $\mathbf{D}^{(1)}$ and (b) $\mathbf{D}^{(2)}$. 
Here, it is assumed that $\varepsilon_1>0$ and $\varepsilon_2>0$.}
 \end{figure}

First, consider weighted charges $q_1 =\epsilon_1(\mathbf{r}_0) q$ and $q_2=\epsilon_2(-\mathbf{r}_0) q$ with $\mathbf{r}_0 = [a\ 0\ 0]^\mathrm{T}$, as shown in Fig.~\ref{fig:charge_response}(a), where the relative permittivity is given as $\epsilon_i = \varepsilon_i/\varepsilon_0$. 
The electric-displacement field at $\mathbf{r}=[x\ y\ z]^\mathrm{T}$ is given as follows:
\begin{equation}
 \mathbf{D}^{(1)}(\mathbf{r})
= 
\begin{cases}
\frac{\epsilon_1(\mathbf{r}_0) q}{4\pi}
\left(\frac{\mathbf{r}-\mathbf{r}_0}{|\mathbf{r}-\mathbf{r}_0|^3} + \frac{\mathbf{r} + \mathbf{r}_0}{|\mathbf{r} + \mathbf{r}_0|^3} \right) & (x\geq 0^+ )\\
\frac{\epsilon_2(-\mathbf{r}_0) q}{4\pi}
\left(\frac{\mathbf{r}-\mathbf{r}_0}{|\mathbf{r}-\mathbf{r}_0|^3} + \frac{\mathbf{r} + \mathbf{r}_0}{|\mathbf{r} + \mathbf{r}_0|^3} \right) & (x\leq 0^- )
\end{cases}   \label{eq:16}
\end{equation}
$\mathbf{D}^{(1)}$ satisfies Maxwell's equations with charge in $x>0$ and $x<0$.
Because $D_x = 0$ holds on $x=0$, the normal component $D_x$ is continuous on $x=0$.
The electric field fulfills the tangential continuity condition on $x=0$,
owing to the appropriate choice of the charge weights with Eq.~(\ref{eq:15}).
Second, we consider $q_1 = q$ and $q_2=-q$, as shown in Fig.~\ref{fig:charge_response}(b).
The electric displacement is given as follows:
\begin{equation}
 \mathbf{D}^{(2)}(\mathbf{r}) = 
\frac{q}{4\pi}\left(\frac{\mathbf{r}-\mathbf{r}_0}{|\mathbf{r}-\mathbf{r}_0|^3} -\frac{\mathbf{r}+\mathbf{r}_0}{|\mathbf{r}+\mathbf{r}_0|^3}\right).  \label{eq:17}
\end{equation}
Equation~(\ref{eq:17}) satisfies Maxwell's equations in $x>0$ and $x<0$ and is antisymmetric with respect to $\mathcal{M}_x$; therefore,
the normal continuity condition of $D_x$ on $x=0$ is satisfied.
The corresponding electric field does not have
a tangential component on $x=0$; hence, 
the tangential continuity condition on $x=0$ is satisfied.

$\mathbf{D}^{(1)}$ and $\mathbf{D}^{(2)}$ represent two solutions for $\epsilon_2(-\mathbf{r}_0) \ne -\epsilon_1(\mathbf{r}_0)$.
By combining Eqs.~(\ref{eq:16}) and (\ref{eq:17}), we can calculate the field for a single charge (e.g.,\ $q_1=q$ and $q_2=0$).
However, $\epsilon_2(-\mathbf{r}_0) = -\epsilon_1(\mathbf{r}_0)$ results in $\mathcal{CM}_x$ symmetry, and the weighted charge distribution for $\mathbf{D}^{(1)}$ becomes \textit{exactly the same} as that for $\mathbf{D}^{(2)}$; i.e.,
the charge distribution does \textit{not} uniquely determine the $\mathcal{CM}_x$ field.
Therefore, $\epsilon_2(-\mathbf{r}_0) = -\epsilon_1(\mathbf{r}_0)$ leads to the singular response for free charge. 
When the limit of $a\rightarrow 0$ is taken, $\mathbf{D}^{(1)}$ and $\mathbf{D}^{(2)}$ provide the $\mathcal{CM}_x$ point and dipole fields, respectively.
Here, we can exclude the source by removing the slab region $\{(x,y,z)|x \in [-a,a],\ y, z \in \mathbb{R}\}$ 
and joining $x=-a-0^+$ and $x=a+0^+$ under $\mathcal{CM}_x$ symmetry.

\subsection{Surface Plasmon Polaritons Originating from Symmetry Protection \label{sec:spp}}

In this section, we show that surface plasmon polaritons originate from 
the symmetry-protected electric zero modes.
Consider a boundary between the uniform permittivities of $\varepsilon_1>0$ in $x\geq 0^+$ and $\varepsilon_2< 0$ in $x\leq 0^-$.
The whole space is assumed to have vacuum permeability $\mu_0$.
We focus on the transverse-magnetic (TM) surface mode with angular frequency $\omega$ and wavenumber $k_y$ along the $y$ direction.
The surface impedance on $x=0$ for $x\geq 0^+$ and $x\leq 0^-$ is denoted as $Z_1$ and $Z_2$, 
respectively, which are expressed as follows:
\begin{equation}
 Z_i = -\jj \frac{\sqrt{{k_y}^2 -\varepsilon_i\mu_0 \omega^2}}{\omega \varepsilon_i}.  \label{eq:18}
\end{equation}
The derivation is summarized in Appendices~\ref{sec:basicEqs}--\ref{sec:uniform_half}.
The resonance condition $Z_1 + Z_2 = 0$, which is equivalent to the continuity conditions of the electric field,
provides the well-known dispersion relation as follows:
\begin{equation}
 k_0 = \frac{\omega}{c_0} = k_y \sqrt{\frac{\epsilon_1+\epsilon_2}{\epsilon_1\epsilon_2}},  \label{eq:19}
\end{equation}
where $c_0=1/\sqrt{\varepsilon_0\mu_0}$, $k_0 = \omega/c_0$, and $\epsilon_i$ represent the speed of light in vacuum, vacuum wavenumber, and
relative permittivity $\epsilon_i = \varepsilon_i/\varepsilon_0$, respectively.

Equation~(\ref{eq:19}) gives the flat zero band for $\epsilon_2 = -\epsilon_1$.
The zero modes accompany $Z_1= -\jj \infty$ and $Z_2 = +\jj \infty$ at $\omega\rightarrow 0^+$, 
indicating that only the electric field appears owing to the electromagnetic decoupling at direct current (DC) limit.
The origin of the zero modes is the $\mathcal{CM}_x$-symmetrized modes shown in Figs.~\ref{fig:CMx_single_layer} or \ref{fig:CMx_double_layer}.
There are degenerated modes located at different positions on $x=0$.
Because these zero modes do not couple with each other, they form the flat zero band.
The eigenfunction with a wavenumber $k_y$ along $y$ is obtained by summing 
$\mathcal{CM}_x$ point or dipole fields
with the weight of the $\exp(-\jj k_y y)$ factor like
$\int_{x=0} \dd S\, \mathbf{E}^{(\mathrm{pt})}_\mathbf{R} \exp(-\jj k_y Y)$, 
where we use $\mathbf{R}=[0\ Y\ Z]^\mathrm{T}$ and
the point charge is replaced with the charge density.
For $k_y\ne 0$, 
$\int_{x=0} \dd S\, \mathbf{E}^{(\mathrm{pt})}_\mathbf{R} \exp(-\jj k_y Y)$
and $\int_{x=0} \dd S\, \mathbf{E}^{(\mathrm{dp})}_\mathbf{R} \exp(-\jj k_y Y)$
yield the same eigenmode, because 
it has both $D_x\ne 0$ and $E_y \ne 0$ components.
From these observations, we can conclude that 
the $\mathcal{CM}_x$ point and dipole fields coalesce.
The square-root function in Eq.~(\ref{eq:19}) is multi-valued in the complex plane. 
This multi-value characteristic indicates that $\epsilon_2 = -\epsilon_1$ is the exceptional point where the two modes typically coalesce \cite{Bergholtz2021}.
For $k_y=0$, point and dipole fields are decoupled, yielding two modes that produce 
two waves localized at $x=0$ and $\infty$.

Now, we can state that $\mathcal{CM}_x$-protected zero modes 
are the origin of surface plasmon polaritons.
Starting from $\epsilon_2 = -\epsilon_1$,
we decrease $\epsilon_2$ to $\epsilon_2 < -\epsilon_1$.
The flat zero band then becomes a finite frequency band.
As the deformation induces unbalanced Poynting vectors in $x\geq 0^+$ and $x \leq 0^-$,
the energy can propagate with a nonzero group velocity. 
This $\mathcal{CM}_x$-broken mode is usually observed 
as surface plasmon polaritons in experiments.

\subsection{Robustness of Symmetry Protection \label{sec:robustness}}
\begin{figure*}[t!]
 \includegraphics{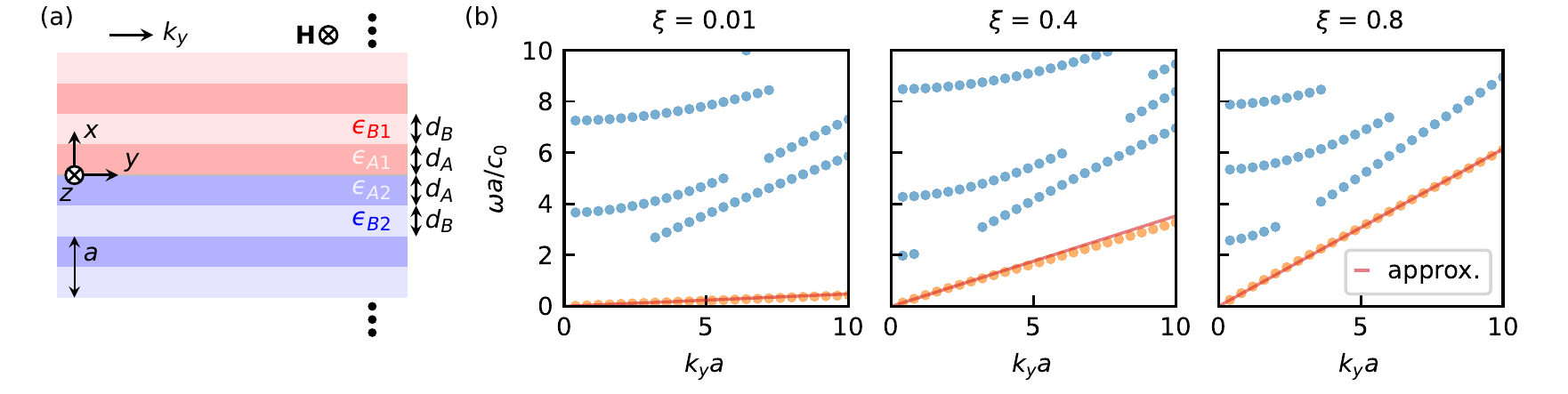}
  \caption{
Surface-wave formation in layered media:
(a) Configuration of dielectric and metallic layers. 
The relative permittivities are set as follows:
$\epsilon_{A1}=(1-\xi)\epsilon_A+\xi$, $\epsilon_{B1}=(1-\xi)\epsilon_B+\xi$,
$\epsilon_{A2}=-\epsilon_A$, and $\epsilon_{B2}=-\epsilon_B$ with
$\epsilon_A=4$ and $\epsilon_B=2$.
The thickness of the layers is $d_A/a = d_B/a=0.5$
with the period $a=d_A+d_B$.
 (b) Calculated TM dispersion relations for $\xi=0.01,\ 0.4,\ 0.8$, depicted as circles.
The lowest band, which is represented by orange circles, originates from $\mathcal{CM}_x$-protected zero modes at $\xi=0$.
The red solid line represents the dispersion calculated via effective-medium approximation, as described in Appendices~\ref{sec:effective_media} and \ref{sec:anisotropic_spp}.
\label{fig:plasmophotonic_layered}}
\end{figure*}

The $\mathcal{CM}_x$-symmetry protection works
for both constant and nonuniform permittivities.
Here, we provide two examples to 
support the robustness of the $\mathcal{CM}_x$ protection for a nonuniform permittivity configuration.

\paragraph{Layered media.}

Consider a layered system with metallic and dielectric materials, as shown in Fig.~\ref{fig:plasmophotonic_layered}(a).
The binary dielectric layers are periodically aligned in $x\geq 0^+$, 
whereas the binary metals are periodically arranged 
with the period $a$ in $x\leq 0^-$.
The thicknesses $d_A$ and $d_B$ of the layers are $a/2$. 
In $x\leq 0^-$, $\epsilon_{A2}=-\epsilon_A$ and $\epsilon_{B2}=-\epsilon_B$ with $\epsilon_A=4$ and $\epsilon_B=2$ are assumed.
Conversely, we set the relative permittivities in $x\geq 0^+$ as 
$\epsilon_{A1}=(1-\xi)\epsilon_A+\xi$ and $\epsilon_{B1}=(1-\xi)\epsilon_B+\xi$, using the parameter $\xi$.
At $\xi = 0$, the $\mathcal{CM}_x$ symmetry holds, whereas $\xi\ne 0$ breaks the symmetry.
We focus on nonradiative localized TM surface waves with wavenumber $k_y$ along $y$.
Let $Z_1$ and $Z_2$ be surface impedances at $x=0$ for $x\geq 0^+$ and $x\leq 0^-$, respectively.
These impedances can be calculated as the Bloch impedances, as described in Appendix~\ref{sec:periodic}.
The surface-wave resonant condition is represented by $Z_1 + Z_2 = 0$.
For a given discretized $k_y$, we numerically evaluate the resonant angular frequency $\omega$
for purely imaginary $Z_1$ and $Z_2$.

Figure~\ref{fig:plasmophotonic_layered}(b) shows
real bands of TM surface waves for $\xi=0.01$, $0.4$, and $0.8$.
The first band (represented by orange circles) distinctly originates 
from the flat zero modes that are $\mathcal{CM}_x$-protected at $\xi=0$.
As $\xi$ increases from 0, the $\mathcal{CM}_x$ symmetry is broken, and 
the first band is raised from zero.
The higher bands are sometimes broken because the wave becomes leaky and propagates into infinity.
Conversely, the plasmonic first band is nonradiative,
and it remains continuous under the perturbation by $\xi$.
This plasmonic dispersion agrees well with 
the solid red line calculated via the effective-medium approximation, as described
in Appendices~\ref{sec:effective_media} and \ref{sec:anisotropic_spp}.

\paragraph{Corrugated system.}

\begin{figure*}[t]
 \includegraphics{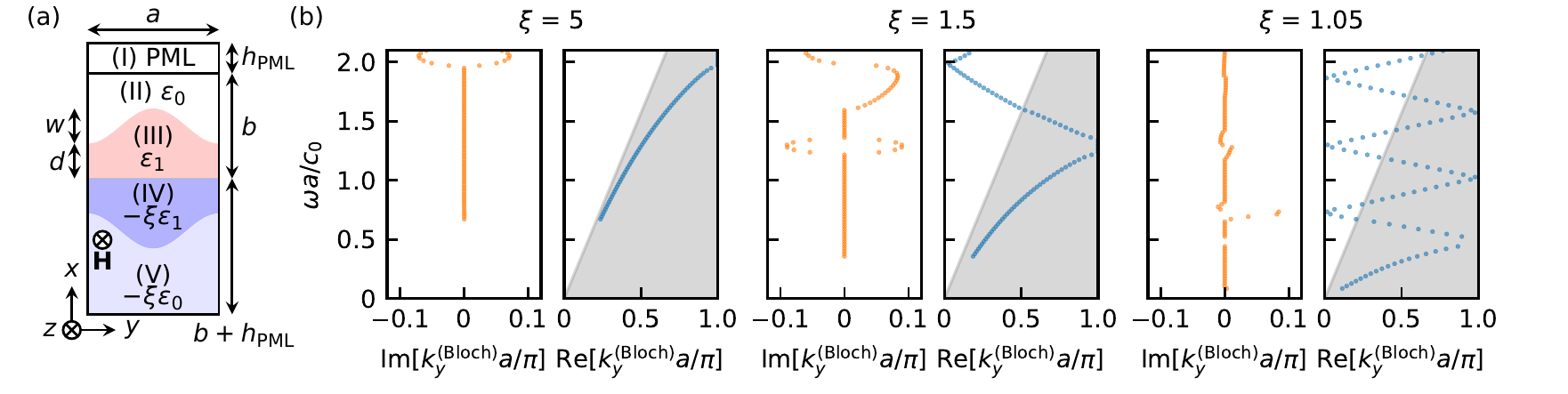}
  \caption{
Surface-wave formation in corrugated system:
(a) Configuration of the simulation. 
I:~PMLs in $b \leq x\leq b+h_\mathrm{PML}$
with the matched permittivity and permeability tensors $\varepsilon_\mathrm{PML}/\varepsilon_0 = \mu_\mathrm{PML}/\mu_0=\mathrm{diag}(\gamma^{-1},\gamma, \gamma)$ with $\gamma = 1-\jj (x-b)^2/L^2$ and $L=0.3 h_\mathrm{PML}$.
II:~vacuum with permittivity $\varepsilon_0$. III:~dielectric material in $0^+ \leq x \leq d + (w/2)[1 - \cos(2\pi y/a)]$ with permittivity $\varepsilon_1 = 5\varepsilon_0$. IV:~metal in $-d - (w/2)[1- \cos(2\pi y/a)] \leq x \leq 0^-$ with permittivity $-\xi\varepsilon_1$. V:~metal with permittivity $-\xi\varepsilon_0$.
The geometric parameters are given as $b/a=6$, $w/a=0.25$, $h_\mathrm{PML}/a=3$, and $d/a=0.1$.
The boundary condition of a perfect electric conductor was imposed at the bottom $x=-(b+h_\mathrm{PML})$ and top $x=b+h_\mathrm{PML}$ boundaries, whereas the sides $y=\pm a/2$ are periodic boundaries.
(b) Complex dispersion relations of TM surface waves for $\xi=5,\ 1.5$, and $1.05$. 
The nonradiative region below the light line is colored gray.
\label{fig:plasmophotonic_corrugated}}
\end{figure*}

Next, we consider nonuniformity in the $y$ direction.
Figure~\ref{fig:plasmophotonic_corrugated}(a) shows 
the unit cell of a corrugated plasmonic/photonic system
with a parameter $\xi$. 
The dielectric--metal boundary is located on $x=0$.
The corrugation is periodic in the $y$ direction 
and is given by the $\cos$ function.
The geometric parameters are set as 
$b/a=6$, $w/a=0.25$, $h_\mathrm{PML}/a=3$, and $d/a=0.1$.
The permittivity in III is set as $\varepsilon_1/\varepsilon_0 = 5$.
At $\xi=1$, the system is $\mathcal{CM}_x$-symmetric.
We calculated the complex Bloch wavenumber $k_y^\mathrm{(Bloch)}(\omega)$ 
of TM surface waves for a given angular frequency $\omega$ 
using the conventional finite-element solver COMSOL Multiphysics \cite{Davanco2007}.
To simplify the plot, 
we restrict $k_y$ satisfying $\operatorname{Re}[k_y^\mathrm{(Bloch)}]\geq 0$.
As the finite-element eigenmode analysis suffers from unphysical modes near the perfectly matched layers (PMLs) \cite{Parisi2012}, we filter the physical modes localized near the surface,
using $\int_\mathrm{I} |\tilde{H}_z|^2\, \dd S/\int_{\mathrm{III}\cup \mathrm{IV}}|\tilde{H}_z|^2 \dd S<0.12$ with the complex amplitude of the $z$ component of the magnetic field $\tilde{H}_z$.
Here, the complex amplitude $\tilde{\mathbf{H}}$ is defined as $\mathbf{H} = \tilde{\mathbf{H}}\exp(\jj \omega t) + \mathrm{c.c.}$ for the real magnetic field $\mathbf{H}$, where $t$ represents time.

Figure~\ref{fig:plasmophotonic_corrugated}(b) presents the calculated complex dispersion relations for $\xi = 5$, $1.5$, and $1.05$.
The wavenumber becomes complex above the light line
because the diffraction by corrugation leads to energy leakage to infinity. 
This observation validates the results.
At $\xi=1$, there is a $\mathcal{CM}_x$-protected zero mode at each point on the dielectric--metal boundary $x=0$.
These degenerated modes produce infinite bands in $\xi>1$.
Therefore, the $\mathcal{CM}_x$-protected zero modes are regarded as the sources of infinite bands.
When $\xi$ approaches 1, 
the frequencies of all plasmonic bands decrease to zero.
Note that lower eigenfrequencies are missing 
due to simulation limitation i.e.,\ these modes are weakly confined and affected by the finite simulation domain.

\begin{figure}[tb]
 \includegraphics{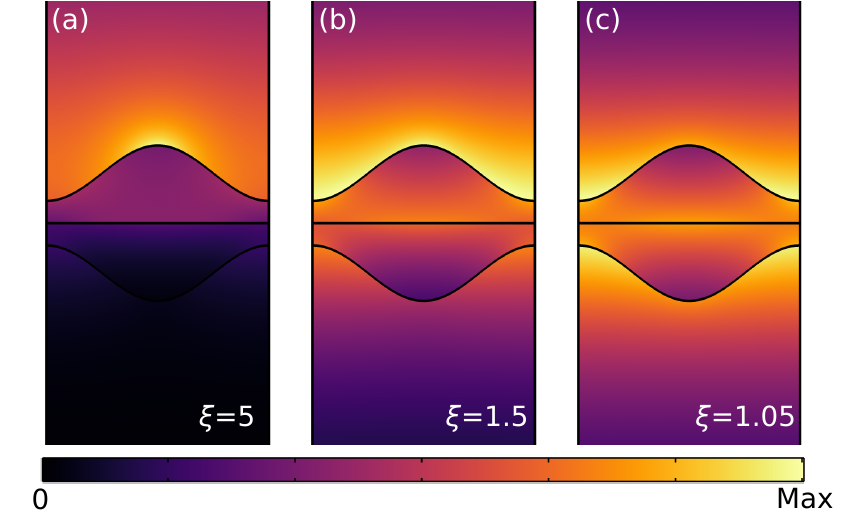}
  \caption{Electric-field amplitude 
$|\tilde{\mathbf{E}}|$ at the first band.
The actual parameters are given as follows:
(a) $\xi=5$, $k_y^\mathrm{(Bloch)} a/\pi =0.495$, $\omega a/c_0 =1.28$,
(b) $\xi=1.5$, $k_y^\mathrm{(Bloch)} a/\pi =0.505$, $\omega a/c_0 =0.838$, 
and (c) $\xi=1.05$, $k_y^\mathrm{(Bloch)} a/\pi =0.500$, $\omega a/c_0 =0.314$.
\label{fig:plasmophotonic_corrugated_fields}
}
\end{figure}

To illustrate the $\mathcal{CM}_x$ protection more explicitly, 
we present the electric-field amplitude $|\tilde{\mathbf{E}}|$ 
of the first band at approximately $k_y^\mathrm{(Bloch)} a/\pi \approx 0.5$ in Fig.~\ref{fig:plasmophotonic_corrugated_fields}.
When $\xi$ approaches 1, the field becomes symmetric with respect to the dielectric--metal boundary $x=0$. This result supports the crucial role of $\mathcal{CM}_x$ symmetry in the formation of surface plasmon polaritons even if the system has nonuniformity.

\subsection{Zero-mode Excitation at Temporal Boundary \label{sec:temporal_boundary}}

We showed that surface plasmon polaritons originate 
from the $\mathcal{CM}_x$-protected zero modes. 
However, realistic configurations break the $\mathcal{CM}_x$ symmetry;
hence, we cannot experimentally observe the surface plasmon polaritons at zero frequency. 
Now, a question arises: can we observe any zero modes originating from a similar symmetry? We answer this question by constructing magnetic zero modes observable at a temporal boundary.

Assume that the permeability satisfies $\mu(x,y,z)>0$ in $x\geq 0^+$.
We consider a localized magnetostatic field in $x\geq 0^+$.
The magnetic field and magnetic flux density in $x\geq 0^+$
are denoted as $\mathbf{H}_0$ and $\mathbf{B}_0$, respectively.
Let $\mathcal{M}_x$ be the mirror-reflection operation with respect to $x=0$.
The $\mathcal{M}_x$-symmetrized magnetic fields $(\mathbf{H}_S, \mathbf{B}_S)$ in the entire space are constructed via the concatenation of 
$(\mathbf{H}_0,\mathbf{B}_0)$ and $\mathcal{M}_x(\mathbf{H}_0,\mathbf{B}_0)$.
We should carefully consider the axial (twisted) characteristics of magnetic fields when we operate $\mathcal{M}_x$.
The symmetrized fields satisfy Maxwell's equations under the $\mathcal{M}_x$-symmetrized permeability $\mu_\mathcal{M}$ is expressed as follows:
\begin{equation}
 \mu_\mathcal{M}(x,y,z) = 
\begin{cases}
 \mu(x,y,z) & (x\geq 0)\\
 \mu(-x,y,z) & (x\leq 0)
\end{cases}   \label{eq:20}
\end{equation}
On $x=0$, the symmetry automatically ensures the continuity condition on $B_x$.
Conversely, tangential magnetic fields ($H_y$ and $H_z$) may not be continuous.
To compensate for the discontinuity, we place a perfect metal at $x=0$ that supports surface currents.

This procedure is then applied to a vacuum.
A magnetic potential $\psi$ is 
introduced to produce a magnetic field $\textbf{H}= -\nabla \psi$.
Let $\tilde{\Psi}_{0}$ be a complex constant and 
consider a magnetic potential $\tilde{\Psi}_{0}\exp(-\jj k_y y) \exp(-k_y x)$
with a wavenumber $k_y$ along the $y$-axis.
The $\mathcal{M}_x$-symmetrized complex magnetic potential omitting $\exp(-\jj k_y y)$ is given as follows:
\begin{equation}
\tilde{\psi}_{S}(x)  =
\begin{cases}
\tilde{\Psi}_{0}\exp(-k_y x) & (x\geq 0^+) \\
-\tilde{\Psi}_{0} \exp(k_y x) & (x\leq 0^-) 
\end{cases}   \label{eq:21}
\end{equation}
Then, the surface current on $x=0$ is given as follows:
\begin{equation}
 \tilde{K}_z(y,z) = 2\tilde{H}_y(0^+, y,z) = 2 \jj k_y \tilde{\Psi}_0 \exp(-\jj k_y y).  \label{eq:22}
\end{equation}
We emphasize that Eqs.~(\ref{eq:21})--(\ref{eq:22}) 
represent a magnetic zero mode,
and the fields are kept unchanged under time evolution.
It is important to see that $\tilde{B}_x(x=0,y,z) = \mu_0 k_y \tilde{\Psi}_0 \exp(-\jj k_y y) \ne 0$ for $k_y \tilde{\Psi}_0 \ne 0$.
Therefore, the magnetic flux perpendicular to the perfect electric conductor on $x=0$ is 
frozen, which is similar to flux pinning in a superconductor.

\begin{figure}[bt]
 \includegraphics{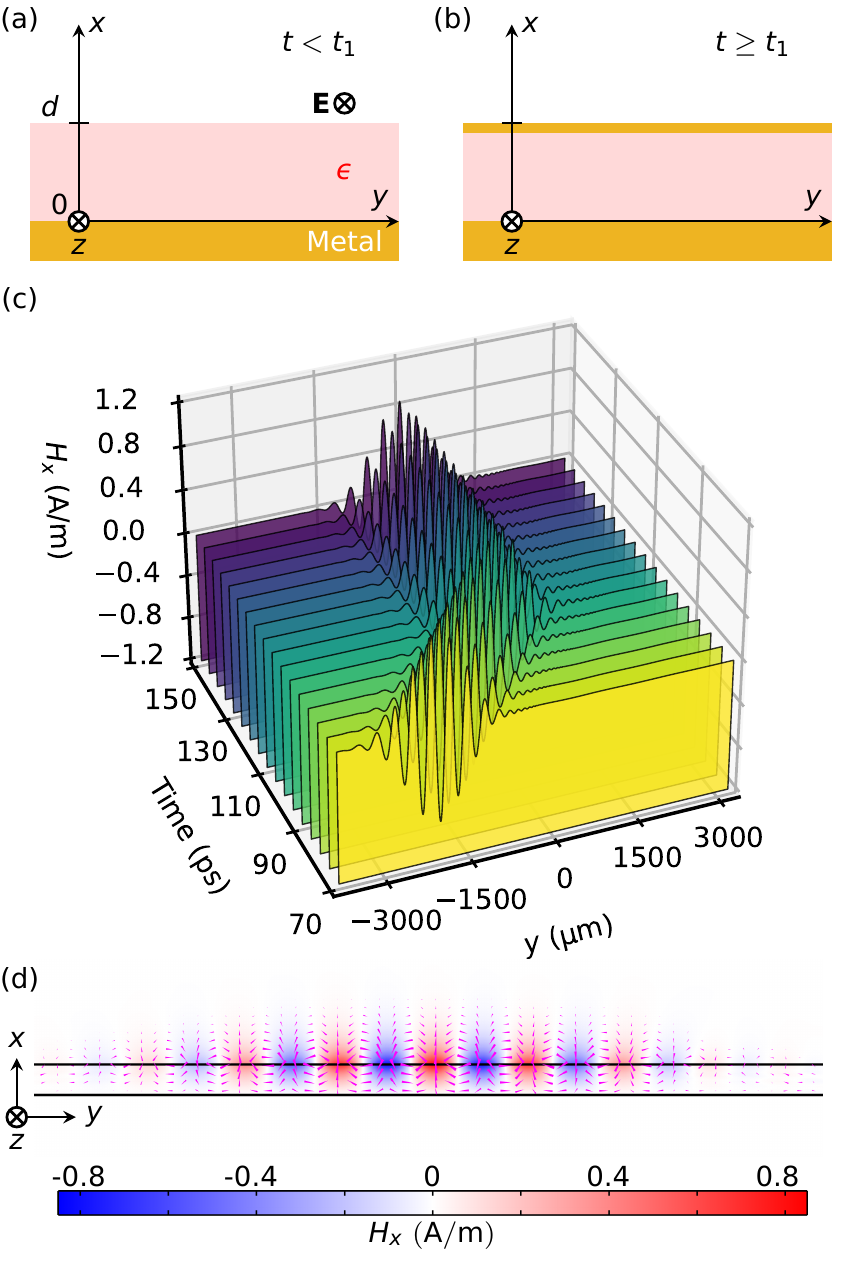}
  \caption{Zero-mode excitation at a temporal boundary:
(a) half-metalized dielectric waveguide; (b) double-metalized dielectric waveguide; (c) temporal dynamics of a magnetic field $H_x$ at the top surface ($x=d$). The waveguide surface is metalized at $t=t_1 = \SI{100}{ps}$. (d) Magnetic-field distribution of the zero mode near $y=0$ at $t=\SI{300}{ps}$. The magenta arrows indicate $\mathbf{H}$, where the arrow length is proportional to the logarithm of the magnitude. The colormap represents $H_x$.
\label{fig:temporal_boundary}}
\end{figure}
Next, we numerically demonstrate the flux pinning at a temporal boundary.
The waveguide is made of a dielectric material with relative permittivity 
$\epsilon$ and occupies $0\leq x\leq d$ in vacuum conditions [Fig.~\ref{fig:temporal_boundary}(a)].
The bottom side ($x=0$) of the dielectric is assumed to be a perfect electric conductor.
Such a half-metalized dielectric waveguide possesses surface modes.
We focus on the transverse-electric (TE) surface modes ($E_x=0$, $E_y=0$) with a wavenumber $k_y$ along the $y$-axis.
Consider a lowest-band TE Gaussian wave packet. 
The wave packet is uniform in the $z$ direction and 
propagates in the $+y$ direction
over time.
At time $t=t_1$, we metalize the top surface ($x=d$) 
and maintain it after $t\geq t_1$ [Fig.~\ref{fig:temporal_boundary}(b)].
Therefore, a temporal boundary appears at $t=t_1$,
resulting in scattering with zero-mode excitation.
The source of the zero mode is the surface current on $x=d$.
The zero mode survives even if the mirror symmetry with respect to $x=d$ is broken.
The above configuration was realized experimentally
for a terahertz wave packet via the photoexcitation of GaAs.
Following Ref.~\onlinecite{Miyamaru2021},
we selected $d=\SI{100}{\micro m}$ and $\epsilon = 12.96$ (GaAs). 
The time-domain dynamics were calculated using COMSOL Multiphysics.
Details and justification of the simulation are presented in Appendix~\ref{sec:details_temporal_boundary}.

Figure~\ref{fig:temporal_boundary}(c) shows the time evolution of $H_x$ on the top surface on $x=d$.
This top surface is metalized at $t_1 = \SI{100}{ps}$.
At $t<t_1$, the wave packet propagates in the $+y$ direction.
For $t\geq t_1$, the wave packet is completely pinned,
considering that the metallic boundary does not carry finite-frequency modes with $H_x\ne 0$ at the boundary.
A movie illustrating the temporal dynamics of the two-dimensional $H_x$ is presented
in Supplemental Material
\footnote{A movie illustrating magnetic-field dynamics near a temporal boundary
is provided as an ancillary file in arXiv.}.
These results indicate the zero-mode excitation at the temporal boundary.
The zero mode was expected to be excited in the experiments of Ref.~\onlinecite{Miyamaru2021},
although the zero-mode observation poses a technical challenge.
Because the mechanism of flux pinning is universal 
and independent of the frequency,
flux pinning can be experimentally observed 
at a frequency below the terahertz range for a thick waveguide.

To examine the topological characteristics of the zero modes, 
we plotted the magnetic-field vector with $H_x$ near $y=0$
at $t=\SI{300}{ps}$, as shown in Fig.~\ref{fig:temporal_boundary}(d).
Interestingly, the magnetic field rotates as $y$ increases.
The rotation direction is determined in each $0<x<d$ and $x>d$.
This feature can be explained as follows:
For $x> d$, a magnetic potential 
$\tilde{\Psi}(x) = \tilde{\Psi}_0\exp[-k_y (x-d)]$ 
produces a circular rotation in the localized magnetic field.
For $0< x< d$, there are right- and left-circular components.
The boundary condition $B_x =0$ on $x=0$ determines the weights of these components. 
Then, right- and left-circular amplitudes equally appear at $x=0$.
Because one of them represents an attenuated solution along $x$, 
the other component is superior.
Therefore, the rotating direction is also determined in $0<x<d$. 
The whole solution is constructed by continuously connecting these solutions at $x=d$.

Finally, we comment on the analogy between electric and magnetic surface plasmon polaritons.
To construct the magnetic analogy to surface plasmon polaritons, 
we introduce the magnetic conjugation operation $\mathcal{C}_m: (\mathbf{H},\mathbf{B})\mapsto (\mathbf{H},-\mathbf{B})$ for a magnetic field $\mathbf{H}$ and magnetic flux density $\mathbf{B}$. The $\mathcal{C}_m \mathcal{M}_x$-protected 
magnetic plasmon polaritons appear at zero frequency, in addition to the electric ones. 
Note that the pinning modes excited at the temporal boundary differ from the magnetic surface plasmon polaritons.
The pinning modes are originally protected by $\mathcal{M}_x$ symmetry rather than
the $\mathcal{C}_m \mathcal{M}_x$ symmetry.

\section{\label{sec:topo_rot} Topological Polarization Rotation and Surface Impedance}

In this section, we examine the general characteristics of the zero-mode field distribution.
The zero modes accompany vector-field rotation, as shown in Fig.~\ref{fig:temporal_boundary}(d).
We analyze a similar rotation in electric zero modes and
identify Keller--Dykhne self-duality as its physical origin.
Lastly, we show that the rotation is directly related to surface impedance,
which characterizes the half-space response.

\subsection{Uniform Layer \label{sec:uniform_layer}}

In this subsection, we investigate polarization rotation 
in a uniform medium
with the Keller--Dykhne self-duality between the electric field and electric displacement.

\paragraph{Solutions.}
The uniform dielectric (or metallic) slab is an elementary building block
for studying surface waves in layered media.
Here, we establish the fundamental property of the basic solutions.
Consider an \textit{electrostatic} field in a slab with 
uniform scalar permittivity $\varepsilon\ne 0$ located in $|x|\leq d/2$ ($d>0$). 
Assume that the wavenumber is given by $k_y > 0$ in the $y$ direction,
and the electric field is on the $xy$ plane.
The boundary conditions on $x=\pm d/2$ are regarded as arbitrary.
The electrostatic potential is represented by
$\tilde{\varphi}(x) \exp(-\jj k_y y)$.
Using constants $\tilde{\Phi}_1$ and $\tilde{\Phi}_2$,
the two independent solutions are expressed as follows:
\begin{align}
 \tilde{\varphi}_1(x) = \tilde{\Phi}_1 \exp\left[-k_y \left(x+\frac{d}{2}\right)\right],  \label{eq:23}\\
\tilde{\varphi}_2(x) = \tilde{\Phi}_2 \exp\left[ k_y \left(x-\frac{d}{2}\right)\right],  \label{eq:24}
\end{align}
which represent the waves localized at $x=-d/2$ and $d/2$, respectively.
Note that a variable with a tilde 
dependent only on $x$ represents the complex amplitude omitting $\exp(-\jj k_y y)$
in our convention for layered media.
The corresponding electric fields are calculated as follows:
\begin{align}
 \tilde{\mathbf{E}}_1(x) &= k_y \tilde{\Phi}_1 \exp\left[-k_y \left(x+\frac{d}{2}\right)\right]
\begin{bmatrix} 
 1\\ \jj \\ 0
\end{bmatrix}
,  \label{eq:25}\\
 \tilde{\mathbf{E}}_2(x) &= -k_y \tilde{\Phi}_2\exp\left[k_y \left(x-\frac{d}{2}\right)\right]
\begin{bmatrix}
 1\\ -\jj \\ 0
\end{bmatrix}.
  \label{eq:26}
\end{align}
These fields involve circular polarizations,
although they do not evolve with time because they are electrostatic.
Considering the omitted $\exp(-\jj k_y y)$ term, 
the electric field rotates as $y$ increases.
We stress that the polarization and momentum (or wavenumber) are locked;
the wavenumber is flipped when we reverse the polarization rotation.
In fact, we can observe 
$k_y \rightarrow - k_y$ and $[1\ \jj\ 0]^\mathrm{T} \leftrightarrow [1\ -\jj\ 0]^\mathrm{T}$
by applying the mirror reflection $\mathcal{M}_y:y\mapsto -y$
for Eqs.~(\ref{eq:25}) and (\ref{eq:26}).

Focusing on the mirror symmetry on $\mathcal{M}_x$, 
we can construct $\mathcal{M}_x$-symmetric and antisymmetric solutions as follows:
\begin{align}
 \tilde{\varphi}_S &= \tilde{\Phi}_S \left(\exp[-k_y (x+d/2)] + \exp[k_y (x-d/2)]\right),  \label{eq:27}\\
 \tilde{\varphi}_A &= \tilde{\Phi}_A \left(\exp[-k_y (x+d/2)] - \exp[k_y (x-d/2)]\right),  \label{eq:28}
\end{align}
where $\tilde{\Phi}_S$ and $\tilde{\Phi}_A$ are constants.
We consider that 
these symmetric and antisymmetric fields are defined 
on the circle $S^1 = \mathbb{R}/\mathbb{Z}d$,
which is equivalent to an interval $[-d/2,d/2]$ when identifying $x=d/2$ with $x=-d/2$.
The single- and double-layer charges at $x = -d/2$ (mod $d$) in $S^1$
give physical sources for Eqs.~(\ref{eq:27}) and (\ref{eq:28}), respectively.
Therefore, the single- and double-layer sources 
produce the two localized modes at $x=-d/2$ and $d/2$.

For $k_y=0$, special care is required,
considering that a constant electric-field potential implies zero electric field.
When we maintain $k_y \tilde{\Phi}_i$ ($i=S, A$) as constant and take the limit of $k_y \rightarrow 0$,
Eqs.~(\ref{eq:27}) and (\ref{eq:28}) give the electric fields of
$\tilde{\mathbf{E}}_S\propto [0\ 1\ 0]^\mathrm{T}$ and
$\tilde{\mathbf{E}}_A\propto [1\ 0\ 0]^\mathrm{T}$, respectively.
Therefore, these constant fields originate from two modes localized at $x=\pm d/2$.

The constant electric fields $[0\ 1\ 0]^\mathrm{T}$ 
and $[1\ 0\ 0]^\mathrm{T}$ 
are directly related to the topology of $S^1$.
At $k_y=0$, the equation has
$\mathcal{M}_y$ symmetry.
Therefore, the electric field is decoupled into
the $x$ and $y$ components.
The former and latter solutions are 
$\mathcal{M}_y$-symmetric and $\mathcal{M}_y$-antisymmetric, respectively. 
We denote
the parities (i.e.,\ eigenvalues) with respect to $\mathcal{M}_x$ and $\mathcal{M}_y$ 
as $\Pi_x$ and $\Pi_y$, respectively.
The \textit{constant} electrostatic fields with $k_y=0$ 
only exist for $(\Pi_x, \Pi_y)=(+1,-1),\ (-1,+1)$.
Moreover, the source at $x = -d/2$ (mod $d$) in $S^1$ vanishes.
Therefore, $[1\ 0\ 0]^\mathrm{T}$ with $(\Pi_x, \Pi_y)=(-1,+1)$
corresponds to the generator of a de Rham cohomology group of $H^1(S^1)$.
Upon rotating $[1\ 0\ 0]^\mathrm{T}$ by $\pi/2$ radians, 
we obtain $[0\ 1\ 0]^\mathrm{T}$ with $(\Pi_x, \Pi_y)=(1,-1)$,
which is considered as the unit normal on $S^1\subset \mathbb{R}^2$.

\paragraph{Keller--Dykhne Duality.}

The appearance of circular polarization in Eqs.~(\ref{eq:25}) and (\ref{eq:26})
can be explained by the Keller--Dykhne self-duality.

First, we introduce Keller--Dykhne duality \cite{Keller1964,Dykhne1971,Mendelson1975,Nakata2019a}.
Let $\mathbf{E}(x,y)$ and $\mathbf{D}(x,y)$ be 
a static two-dimensional electric field 
and an electric displacement, respectively.
The Keller--Dykhne duality relates $(\mathbf{E},\mathbf{D})$
with its dual $(\mathbf{E}^\star,\mathbf{D}^\star)$.
We assume that these vector fields only have in-plane components of $x$ and $y$
and satisfy Maxwell's equations as follows:
\begin{equation}
 \nabla\cdot \mathbf{D} =0,\quad \nabla \times \mathbf{E} = 0.  \label{eq:29}
\end{equation}
The constitutive relation is expressed as follows:
\begin{equation}
 \mathbf{D}(x,y) = \varepsilon(x,y) \mathbf{E}(x,y).  \label{eq:30}
\end{equation}
Consider that the dual fields defined as follows:
\begin{equation}
\mathbf{E}^\star = {\varepsilon_\mathrm{ref}}^{-1} \mathbf{e}_z \times \mathbf{D},\quad  \mathbf{D}^\star = \varepsilon_\mathrm{ref}  \mathbf{e}_z \times \mathbf{E},  \label{eq:31}
\end{equation}
where $\varepsilon_\mathrm{ref}$ represents the constant permittivity,
and $\mathbf{e}_z$ is the unit vector along the $z$-axis. 
The operation $\mathbf{e}_z \times $ induces rotation by $\pi/2$ radians with respect to the $z$-axis.
These fields satisfy Maxwell's equations, as follows:
\begin{equation}
 \nabla\cdot \mathbf{D}^\star =0,\quad \nabla \times \mathbf{E}^\star = 0.  \label{eq:32}
\end{equation}
The relationship between $\mathbf{D}^\star$ and $\mathbf{E}^\star$ is given as follows:
\begin{equation}
 \mathbf{D}^\star(x,y) = \varepsilon^\star(x,y) \mathbf{E}^\star(x,y),\quad
 \varepsilon^\star(x,y) = \frac{{\varepsilon_\mathrm{ref}}^2}{\varepsilon(x,y)}.  \label{eq:33}
\end{equation}
In summary, the Keller--Dykhne duality can relate solutions in
different permittivity distributions in Eqs.~(\ref{eq:30}) and (\ref{eq:33}).

The uniform slab with constant $\varepsilon$ is \textit{self-dual}
when we choose $\varepsilon_\mathrm{ref}=\varepsilon$.
The solutions can be classified as eigenstates under
the rotation $\mathcal{R}_{\pi/2} = \mathbf{e}_z \times $.
The eigenvectors of $\mathcal{R}_{\pi/2}$ are given as
circular polarizations $\mathbf{e}_x \pm \jj \mathbf{e}_y$ 
with the unit vector $\mathbf{e}_i$ along the $i=x, y$ axes, whereas
the corresponding eigenvalues are $\mp \jj$, respectively.
Therefore, Eqs.~(\ref{eq:25}) and (\ref{eq:26}) involve circular polarizations.

Duality transformation $\mathcal{D}: (\mathbf{E},\mathbf{D})\mapsto (\mathbf{E}^\star,\mathbf{D}^\star)$ does not commute with $\mathcal{M}_x$.
Therefore, $\mathcal{D}$ and $\mathcal{M}_x$ cannot be simultaneously diagonalized.
$\mathcal{D}$ connects $\mathcal{M}_x$-symmetric and $\mathcal{M}_x$-antisymmetric solutions.
In fact, the electric fields obtained from Eqs.~(\ref{eq:27}) and (\ref{eq:28})
are interchanged when the duality transformation of Eq.~(\ref{eq:31}) is applied.

\paragraph{F$_0$ Matrix.}

The conventional F matrix is defined for a pair of electric and magnetic fields, as indicated by Eq.~(\ref{eq:68}).
However, electric and magnetic fields are decoupled at zero frequency.
Thus, 
we introduce an F$_0$ matrix in the electrostatics, by which electric and electric displacement fields are multiplied.
We consider a slab located in $|x|\leq d/2$ with a uniform scalar permittivity $\varepsilon$.
The F$_0$ matrix connects the fields at $x=\pm d/2$ as follows:
\begin{equation}
\begin{bmatrix}
\tilde{D}_x\left(-\frac{d}{2}\right) \\
\tilde{E}_y\left(-\frac{d}{2}\right)
\end{bmatrix} 
=
F_0
\begin{bmatrix}
\tilde{D}_x\left(\frac{d}{2}\right) \\
\tilde{E}_y\left(\frac{d}{2}\right)
\end{bmatrix}  \label{eq:34}
\end{equation}
Using the linear combination of Eqs.~(\ref{eq:25}) and (\ref{eq:26}),
$F_0$ is calculated as follows:
\begin{equation}
F_0 = 
\begin{bmatrix}
 \cosh(k_y d) & - \jj \varepsilon \sinh(k_y d)\\
\jj \varepsilon^{-1} \sinh(k_y d) &  \cosh(k_y d)
\end{bmatrix}  \label{eq:35}
\end{equation}
Here, $\det F_0 = 1$ holds owing to the reciprocity of the scalar permittivity,
as discussed in Appendix~\ref{sec:reciprocity}.

\subsection{Nonuniform Multilayer \label{sec:nonuniform}}

In this subsection, 
we show that the direction of electrostatic polarization rotation is conserved
even in a multilayer dielectric material; hence, it is considered a topological property. 
Next, we examine the relationship between the polarization rotation and
the surface impedance of a half-space.

\paragraph{Multilayer Solution.}

We select $x_0=0< x_1 < x_2< \cdots < x_n$ along $x$.
Each region of $x\in [x_{i-1},x_i]$ is occupied by a uniform slab with the scalar permittivity $\varepsilon_i$ ($i=1,2,\cdots, n$).
The width of each slab is expressed as
$d_i = x_i - x_{i-1}$.
The magnetic permeability is $\mu_0$ for all regions.
We assume that the region 
of $x> x_n$ has a constant permittivity $\varepsilon_{n+1}$.
If $\varepsilon$ is \textit{finite} at zero frequency,
$\tilde{H}_z=0$ holds from Eq.~(\ref{eq:61}).
As the field should not diverge at $x\rightarrow +\infty$,
we may set the field at $x= x_n$ for $k_y>0$ as follows:
\begin{equation}
 \begin{bmatrix}
  \tilde{D}_x(x_n)\\
  \tilde{E}_y(x_n)
 \end{bmatrix}
=
 \begin{bmatrix}
  \varepsilon_{n+1} \\
  \jj
 \end{bmatrix}.  \label{eq:36}
\end{equation}
The continuity condition on $\tilde{D}_x$ and $\tilde{E}_y$
allows us to multiply F$_0$ matrices to obtain the solution.
In fact, we can calculate 
the field at $x$ satisfying $x_{i-1} \leq x < x_{i}$ as follows:
\begin{multline}
 \begin{bmatrix}
  \tilde{D}_x(x)\\
  \tilde{E}_y(x)
 \end{bmatrix}
= F_0(\varepsilon_i, x_i-x )\times\\
 F_0(\varepsilon_{i+1}, d_{i+1}) F_0(\varepsilon_{i+2}, d_{i+2}) \cdots
F_0(\varepsilon_n, d_n)
 \begin{bmatrix}
  \varepsilon_{n+1} \\
  \jj
 \end{bmatrix},  \label{eq:37}
\end{multline}
where the F$_0$ matrix is denoted as
$F_0 (\varepsilon, d)$ with the parameters $\varepsilon$ and $d$.

Even for a continuous distribution of $\varepsilon(x)$, 
we can evaluate Eq.~(\ref{eq:37}) 
by dividing the section into sufficiently small pieces.
Additionally, we can derive a formal solution for
the continuous $\varepsilon(x)$, as discussed in Appendix~\ref{sec:formal}.

\paragraph{Topological Polarization Rotation.}

The multilayer solution of Eq.~(\ref{eq:37}) 
generally includes both left and right circular polarization
owing to self-duality breaking caused by the nonuniform permittivity.
However, when the signature of $ \varepsilon_i$ is the same for all $i$, the direction of polarization rotation is \textit{conserved} in $x\geq 0^+$.

Consider $\varepsilon_i>0$ for all $i=1,2,\cdots, n+1$.
Assume that $[\tilde{D}_x(x_{i})\ \tilde{E}_y(x_{i})]^\mathrm{T}$
has the form of $[\varepsilon_C\ \jj \xi]^\mathrm{T}$ ($\varepsilon_C>0$, $\xi>0$),
which is satisfied by Eq.~(\ref{eq:36}). 
Using the F$_0$ matrix, $[\tilde{D}_x(x_{i-1})\ \tilde{E}_y(x_{i-1})]^\mathrm{T}$ can be calculated as follows:
\begin{equation}
 \begin{bmatrix}
  \tilde{D}_x(x_{i-1})\\ \tilde{E}_y(x_{i-1})
 \end{bmatrix}
=
 \begin{bmatrix}
 \cosh(k_y d_i) \varepsilon_C + \xi\varepsilon_i\sinh(k_y d_i)\\
\jj \left( \frac{\varepsilon_C}{\varepsilon_i}\sinh(k_y d_i) + \xi\cosh(k_y d_i) \right)
 \end{bmatrix}.  \label{eq:38}
\end{equation}
 $[\tilde{D}_x(x_{i-1})\ \tilde{E}_y(x_{i-1})]^\mathrm{T}$ has the same form of $[\varepsilon_C\ \jj \xi]^\mathrm{T}$ ($\varepsilon_C>0$, $\xi>0$), because $k_y d_i$, $\varepsilon_C$, $\varepsilon_i$, and $\xi$ are positive.
Therefore, we deduce that $-\jj \tilde{E}_y(x)/ \tilde{E}_x(x) > 0$ at \textit{any} point in $x\geq 0^+$.
The same conservation law can be justified even for a periodic system with infinite layers [e.g.,\ $x\geq 0^+$ of Fig.~\ref{fig:plasmophotonic_layered}(a)], as shown in Appendix~\ref{sec:prf_tp_rot_per}.

The electrostatic potential $\tilde{\varphi}$ for
$\varepsilon_i > 0$ can also give 
the solution for the permittivity distribution $-\varepsilon_i$.
Therefore, $-\jj \tilde{E}_y(x)/ \tilde{E}_x(x) > 0$ still 
holds at \textit{any} point in $x\geq 0^+$
for $\varepsilon_i<0$ ($i=1,2,\cdots$).

\paragraph{Relationship Between Polarization Rotation and Surface Impedance.}

To relate the polarization rotation to the surface property, 
we consider the case of a finite angular frequency $\omega$.
We define $\Theta = - \jj \tilde{E}_y(x)/\tilde{E}_x(x)$ at $\omega$.
From Eq.~(\ref{eq:61}), $\Theta$ is related to the surface impedance $Z_S = \tilde{E}_y(0)/\tilde{H}_z(0)$ for $x\geq 0^+$
as follows: 
\begin{equation}
Z_S(\omega, k_y) = -\jj \frac{k_y}{\omega \varepsilon_1} \Theta(x=0^+,\omega, k_y),  \label{eq:39}
\end{equation}
where $\varepsilon_1$ represents the permittivity at $x=0^+$, and 
we explicitly express the dependence on $x$, $\omega$, and $k_y$ in $\Theta(x,\omega, k_y)$.

As indicated by the previous discussion on the topological polarization rotation, 
 $\lim_{\omega\rightarrow 0^+} \Theta > 0$ holds.
Therefore, all-positive and all-negative permittivity distributions
lead to
$\lim_{\omega\rightarrow 0^+} \operatorname{Im} Z_S < 0$ and
$\lim_{\omega\rightarrow 0^+} \operatorname{Im} Z_S > 0$, which indicate that the half-space is
capacitive and inductive at zero frequency, respectively.
When $\Theta$ is finite at the DC limit,
we obtain $\lim_{\omega\rightarrow 0^+} |Z_S|=+\infty$,
which indicates the electric--magnetic decoupling at the DC limit.

\section{Bulk--Edge Correspondence to Guarantee Existence of Surface Plasmon Polaritons \label{sec:bulk-edge}}

In this section, we establish bulk--edge correspondence, 
which generally ensures the existence of surface plasmon polaritons even with nonuniformity.

In the previous section, it was shown that 
$W = \lim_{\omega\rightarrow 0^+} \operatorname{sign}(\operatorname{Im}Z_S)$ 
defines a topological quantity in the half-space.
$W$ is directly related to the topological polarization rotation of electrostatic fields.
$W=-1$ and $W=+1$ indicate that the half-space 
exhibits capacitive and inductive behavior, respectively, in the DC limit. 
Now, we conjecture the bulk--edge correspondence as follows:
There always exists a surface mode between two half-spaces with different $W=\pm 1$ for a given $k_y$.
First, we examine examples of the bulk--edge correspondence.
However, it is difficult to justify the bulk--edge correspondence while focusing on static electric fields alone.
To overcome this problem, we consider the magnetic response and frequency dispersion.
Then, the bulk--edge correspondence is generally proved with the circuit-theoretical consideration.

\subsection{Empirical Reasoning \label{sec:emprical}}

In this subsection, we consider examples of 
the surface-wave formation on the boundary between $W=\pm 1$
media to justify the bulk--edge correspondence. 

Consider the boundary between 
a dielectric material and metal, as discussed in Sec.~\ref{sec:spp}.
The dielectric and metal regions with $\varepsilon_1>0$
and $\varepsilon_2<0$
have $W=-1$ and $W=1$, respectively.
Note that the definitions of $\Theta$ and $Z_S$ are changed to
$\Theta = + \jj \tilde{E}_y/\tilde{E}_x$ and 
$Z_S = -\tilde{E}_y(0)/\tilde{H}_z(0)$ for $x\leq 0^-$.
If $\varepsilon_2\leq -\varepsilon_1$, 
Eq.~(\ref{eq:19}) gives a real eigenfrequency. 
Conversely, the imaginary eigenfrequency appears
if $-\varepsilon_1 < \varepsilon_2 < 0$ is satisfied;
however, the modes are still localized at $x=0$.
In both cases, there is a surface plasmon polariton for a given $k_y$.

Next, we consider a nonuniform scalar permittivity distribution.
In $x\geq 0^+$, assume that $\varepsilon(x)>0$, which is periodic in $x$ with a period of $a_1$.
In $x\leq 0^-$, $\varepsilon(x)$ satisfies $\varepsilon(x) < 0$, 
which is periodic in $x$ with a period of $a_2$.
Although it is difficult to prove that the boundary has a surface mode generally, 
we can analyze its existence in two specific cases: (i) $k_y a_i \gg 0$ 
and (ii) $k_y a_i \approx 0^+$ for $i=1,2$.
In (i), the surface wave is tightly localized at $x=0$.
Therefore, the configuration is approximated as the boundary between $\varepsilon(0^+)$ and 
$\varepsilon(0^-)$, which is reduced to the previous configuration.
Therefore, the surface wave exists for a given $k_y$.
In (ii), the surface wave is loosely trapped at $x=0$; therefore, we use the effective-medium approximation.
In $x\geq 0^+$, the \textit{effective} relative anisotropic permittivities along the $x$ and $y$ directions are
given as $\epsilon^{(1)}_x$ and $\epsilon^{(1)}_y$, respectively.
Similarly, $\epsilon^{(2)}_x$ and $\epsilon^{(2)}_y$ 
are defined for $x\leq 0^-$.
The surface impedances on $x=0$ are 
denoted as $Z_1$ and $Z_2$ for $x\geq 0^+$ and $x\leq 0^-$, respectively. 
As described in Appendices~\ref{sec:effective_media} and \ref{sec:anisotropic_spp}, 
the resonance condition of $Z_1 +Z_2=0$
gives the following dispersion relation:
\begin{equation}
 k_0 = \frac{\omega}{c_0}=k_y \sqrt{
\frac{\epsilon_x^{(2)}\epsilon_y^{(2)} - \epsilon_x^{(1)}\epsilon_y^{(1)}}{\epsilon_x^{(1)}\epsilon_x^{(2)} \left(\epsilon_y^{(2)}-\epsilon_y^{(1)}\right) }}.  \label{eq:40}
\end{equation}
Additionally, we can directly show that the mode with 
Eq.~(\ref{eq:40}) is bounded on the surface 
and that the field varies slowly in $x$ compared with $a_i$.
Therefore, the surface mode always exists for $k_y a_i \approx 0^+$ ($i=1,2$).
Although the case of a general wavenumber [excluding (i) and (ii)] is difficult to handle rigorously, 
we can heuristically justify the bulk--edge correspondence as follows.
Consider $\varepsilon(x)>0 $ ($x\geq 0^+$) and $\varepsilon(x) <0$ ($x \leq 0^-$).
The magnetic permeability is given by the vacuum permeability $\mu_0$.
Next, we continuously deform $\varepsilon(x)$ in $x\geq 0^+$ so that 
$\mathcal{CM}_x$ symmetry is satisfied, while keeping $W$ constant.
Then, the $\mathcal{CM}_x$ symmetry ensures the existence of the surface zero modes.
Under the deformation, the eigenfrequency can continuously change; that is, 
a new mode is not created, an existing mode is not annihilated,
and a localized state does not suddenly change to a diverged one and vice versa.
We follow the above process in reverse.
Then, the surface mode exists in the original 
configuration as long as its frequency is kept low to avoid energy leakage to infinity.
However, the above reasoning has not been validated.

\subsection{Bulk--Edge Correspondence from Circuit-Theoretical Consideration \label{sec:bulk-edge_from_circuit}}

To overcome the challenge for proving the bulk--edge correspondence, we introduce circuit-theoretical concepts
and use them to prove the bulk--edge correspondence.

\paragraph{Classification of Response at Zero Frequency.}

The passive driving impedance $Z$ must be a positive-real function of
$s= \jj \omega$, where $\omega$ represents the angular frequency \cite{Guillemin1957}.
If the circuit is lossless, $Z(s)$ is an odd function owing to the time-reversal symmetry.
Then, a physically possible lossless impedance is an odd positive-real function
that satisfies $\operatorname{Im}Z|_{\omega= 0} = 0$ or
$\operatorname{Im}Z|_{\omega= 0^+} = -\infty$ \cite{Wing2010}.
We can infer that these responses originate from electric and magnetic zero modes, as discussed in Sec.~\ref{sec:em_zero_modes}.
The reactance theorem ensures that $\operatorname{Im}Z$ increases monotonically
as $\omega$ increases.
Therefore, $\operatorname{Im}Z|_{\omega= 0} = 0$ 
indicates that $Z$ behaves inductively near $\omega=0$.
By contrast, $\operatorname{Im}Z|_{\omega= 0^+} = -\infty$ suggests a capacitive response near $\omega=0$ with $W=-1$.
In summary, the frequency response is classified as $W=\pm 1$ when there is a gap near zero frequency.
The above properties are valid even for a continuous (distributed-element) model because the system can be modeled by finite numbers of inductors and capacitors when we set the spatial discretization small enough for the typical length scale of the focusing phenomena.

The above properties can be confirmed in simple examples.
The first example is a vacuum.
From Eq.~(\ref{eq:18}), the TM vacuum surface impedance is given as follows:
\begin{equation}
 Z_\mathrm{vac} = - \jj \frac{\sqrt{{k_y}^2 -\varepsilon_0 \mu_0 \omega^2}}{\omega\varepsilon_0}.  \label{eq:41}
\end{equation}
Clearly, $\operatorname{Im} Z_\mathrm{vac}|_{\omega= 0^+} = -\infty$ holds, 
and the vacuum is capacitive near zero frequency.
The second example is metal, which can be modelized using the following Drude permittivity \cite{ashcroft1976solid}:
\begin{equation}
 \varepsilon_m = \varepsilon_0 \left[1 - \left(\frac{\omega_p}{\omega}\right)^2\right],  \label{eq:42}
\end{equation}
where $\omega_p$ represents the plasma angular frequency.
Accordingly, the TM surface impedance of metal is expressed as follows:
\begin{equation}
  Z_m = -\jj \frac{\sqrt{{k_y}^2 -\varepsilon_m\mu_0 \omega^2}}{\omega \varepsilon_m}.  \label{eq:43}
\end{equation}
For $\omega \sim 0$, $Z_m$ can be approximated as follows:
$Z_m \simeq \jj (\omega/\omega_p) Z_0 \sqrt{1+ (c_0 k _y/\omega_p)^2}$,
where $Z_0 = \sqrt{\mu_0/\varepsilon_0}$ represents the vacuum impedance.
Then, the metallic half-space behaves as an inductor.
We stress that an electric field does \textit{not} exist (i.e.,\ $\tilde{E}_x=\tilde{E}_y=0$), because $\varepsilon_m$ diverges at zero frequency.
From $Z_m \rightarrow 0$ under $\omega \rightarrow 0$,
only a magnetic field ($\tilde{H}_z \ne 0$)
 can exist.

\paragraph{General Properties of Surface Impedances.}

Consider a dispersive metal with $\varepsilon(\omega, x)< 0$ and $\mu_0$ 
located in $x\leq 0^-$.
For example, we assume
\begin{equation}
 \varepsilon(\omega, x) = \varepsilon_\mathrm{BG}(x) \left[1- \left(\frac{\omega_p(x)}{\omega}\right)^2\right]  \label{eq:44}
\end{equation}
with background permittivity $\varepsilon_\mathrm{BG}(x)>0$ and plasma angular frequency $\omega_p(x)>0$.
Inside the Drude metal, electric fields vanish
at $\omega\rightarrow 0$.
In fact, Eq.~(\ref{eq:61}) is approximated as $\omega k_y \tilde{H}_z = \varepsilon_0 {\omega_p}^2 \tilde{E}_x$
near $\omega=0$.
Therefore, $\tilde{E}_x=0$ holds true and $\tilde{H}_z\ne 0$ is possible at $\omega=0$.
Similarly, $\tilde{E}_y=0$ is deduced from Eq.~(\ref{eq:62}) 
under the assumption of a finite $\dd\tilde{H}_z/\dd x$.
Then, $\operatorname{Im}Z_2|_{\omega= 0} = 0$ is 
expected as the TM metal response.

In $x\geq 0^+$, we consider a distributed $\varepsilon(x)$ with $\mu_0$.
Assume that $\varepsilon(x)$ satisfies $\varepsilon(x)\geq \varepsilon_0$ and 
$\varepsilon(x)\rightarrow \varepsilon_0$ ($x\rightarrow \infty$). 
The first assumption is reasonable because $\varepsilon < \varepsilon_0$ 
usually accompanies strong frequency dispersion.
Additionally, the second assumption is justified, because 
finite region is enough to be considered for localized waves.
Let $Z_1$ be the surface impedance of the half-space $x\geq 0^+$.
At $\omega\rightarrow 0^+$, $\operatorname{Im} Z_1 \rightarrow -\jj \infty$ holds
from the discussion of topological polarization rotation.

Now, we prove that $Z_1$ includes a zero in $0<\omega\leq c_0 k_y$.
This lemma is used in the proof of the bulk–edge correspondence.
We begin with a uniform vacuum. 
The surface impedance of Eq.~(\ref{eq:41}) is purely imaginary,
and $Z_\mathrm{vac}$ has zero at $\omega = c_0 k_y$.
We prove that the zero is kept in $0< \omega \leq c_0k_y$
even when we gradually insert dielectric slabs in $x\geq 0$.
First, we consider adding a dielectric slab 
with thickness $d$ and permittivity $\varepsilon\geq\varepsilon_0$ ($\epsilon=\varepsilon/\varepsilon_0$). This slab is placed in $x\in [-d,0]$.
Second, the whole system is translated by $d$ along $x$
so that the surface is located on $x=0$.
After the insertion and translation, $Z_1$ is still purely imaginary for $\omega \leq c_0 k_y$.
At $\omega=c_0k_y$, the wavenumber along $x$ and impedance inside the added dielectric should satisfy
$k_x =\sqrt{\epsilon {k_0}^2 – {k_y}^2}\geq 0$ and $Z=k_x/(\omega\varepsilon)\geq 0$, respectively.
If there is a zero at $\omega=c_0k_y$ in the initial $Z_1$, 
the insertion and translation lead to $Z_1|_{\omega=c_0 k_y} = \jj Z \tan (k_x d)$ 
as observed from Eq.~(\ref{eq:69}). 
Considering a small $d$, we determine
$\operatorname{Im} Z_1|_{\omega=c_0 k_y} \simeq Zk_x d \geq 0$.
Therefore, the zero at $c_0 k_y$ always goes to the lower frequency after the gradual insertion and translation.
The insertion is repeated until $\varepsilon (x)$ is obtained.
From the above discussion, the gradual insertion of the dielectric keeps the zero inside $(0,c_0 k_y]$.

\paragraph{Proof.}
Now, we complete the proof of the bulk--edge correspondence.
The lowest angular frequency satisfying $Z_1=0$ 
inside $0<\omega \leq c_0 k_y$ is denoted as $\omega=\omega_0$.
First, we consider the case of $\operatorname{Im} Z_2|_{\omega=\omega_0}\geq 0$.
The following three conditions are satisfied:
(i) $Z_1 + Z_2$ is purely imaginary in $0< \omega \leq c_0 k_y$,
and $\operatorname{Im}(Z_1 + Z_2)$ monotonically increases with $\omega$;
(ii) $\lim_{\omega\rightarrow 0^+}\operatorname{Im}(Z_1 + Z_2)=-\infty$; 
and (iii) $\operatorname{Im}(Z_1 + Z_2)|_{\omega=\omega_0}\geq 0$.
Under these conditions, the intermediate value theorem
ensures that there is an angular frequency 
$\omega \in (0, c_0 k_y]$ at which $Z_1 + Z_2 =0$ is satisfied.
There may be poles of $\operatorname{Im}(Z_1 + Z_2) = \pm \infty$, and the intermediate value theorem should be extended to handle infinity.
Even if $\operatorname{Im} Z_2|_{\omega=\omega_0}< 0$ unexpectedly holds,
we can introduce $\omega_1$ as the lowest angular frequency in $(0, \omega_0]$, satisfying
$Z_2 =\pm \jj\infty$ at $\omega_1$.
Then, we can apply a similar discussion for $(0,\omega_1]$ to ensure the existence of a zero in $Z_1+Z_2$.
Thus, there always exists a surface localized mode at the boundary 
between $W=\pm 1$ media, under the physical assumption.

\subsection{Relationship Between Symmetry Protection and Bulk--Edge Correspondence}
Finally, we discuss the relationship between the $\mathcal{CM}_x$-protected zero modes and the
bulk--edge correspondence according to physical constraints.
Although their theoretical foundations differ significantly, they are related to each other.
We examine the relationship while focusing on a conventional surface-plasmon polariton.
Assuming that $x\geq 0^+$ is filled with a dielectric material with constant permittivity $\varepsilon_1>0$ whereas $x\leq 0^-$ is occupied by a metal with the Drude response [Eq.~(\ref{eq:42})], our bulk--edge correspondence ensures the existence of a surface wave
with the dispersion relation $\omega(k_y)$. For a given $k_{y0}>0$, 
we define a frequency-independent permittivity 
$\bar{\varepsilon}_2 = \varepsilon_m|_{\omega=\omega(k_{y0})}<0$.
Now, consider the boundary between $\varepsilon_1$ and $\bar{\varepsilon}_2$.
It gives the original surface plasmon polariton \textit{only}
at $(k_y, \omega) = (k_{y0}, \omega(k_{y0}))$.
When we gradually transform $\bar{\varepsilon}_2$ into $-\varepsilon_1$,
the surface mode at $ (k_y, \omega) = (k_{y0}, \omega(k_{y0}))$ 
becomes the $\mathcal{CM}_x$-symmetric zero mode. 
Thus, the bulk--edge correspondence predicts a mode 
originating from a $\mathcal{CM}_x$-protected zero mode.

\section{Essential Understanding of Dielectric--Metal Transition
Based on Minimal Circuit Model \label{sec:dielectric_metal}}

In Sec.~\ref{sec:bulk-edge}, we established the bulk--edge correspondence, which generally 
explains the existence of surface plasmon polaritons between a metal and a dielectric material
even with nonuniformity.
Now, a question arises: how can we understand the essential difference between metals and dielectrics
from the perspective of the band theory?
Although we can attempt to solve this problem directly using a continuous model,
its complexity blurs the underlying physics.
Therefore, we take a different approach.
We propose and analyze a minimal circuit model, 
which induces a topological transition.
Furthermore, we show that the minimal model can accurately explain
the dielectric--metal transition.
Finally, the transition is understood as the interchange between electric and magnetic zero modes.
Our approach highlights essential physics in the topological transition without distractions.

\subsection{Composite Right/Left-Handed (CRLH) Transmission Line}

The CRLH transmission line was historically introduced 
to investigate the negative refractive index 
from a circuit-theoretical perspective \cite{Caloz2006}.
Herein, we show 
that the CRLH transmission line is considered a minimal model for inducing a topological transition with duality.

\paragraph{Model.}
\begin{figure}[!tb]
 \centering
  \includegraphics{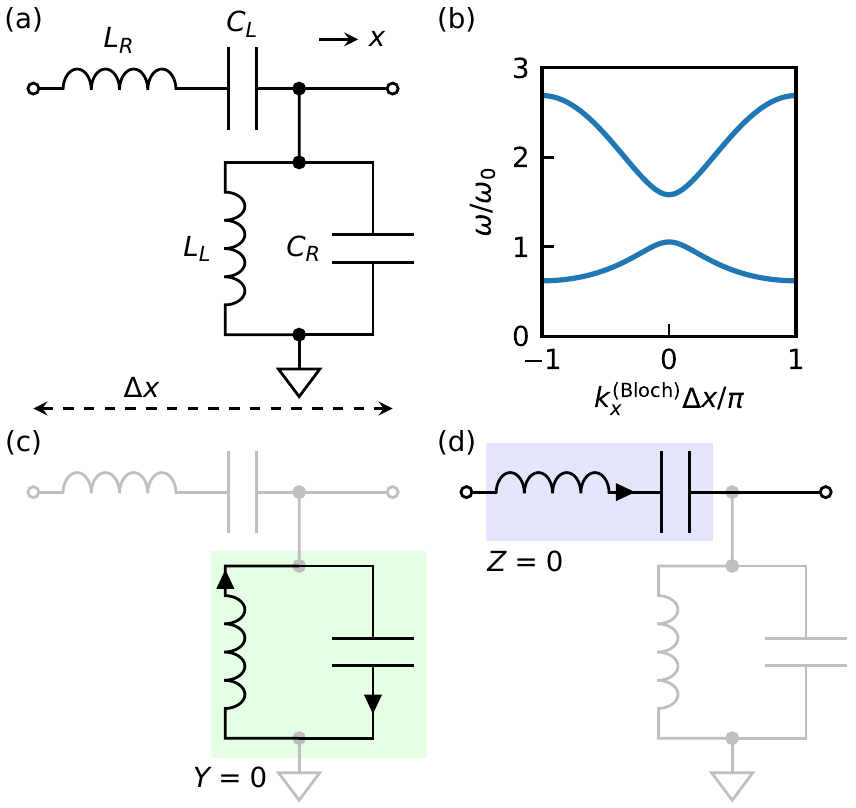}
  \caption{
\label{fig:CRLH}
CRLH transmission line:
(a) Unit cell with period $\Delta x$.
The subscripts $L$ and $R$ denote the left- and right-handed components, respectively.
(b) Typical dispersion relation of a CRLH transmission line. 
Here, we define $\omega_0 = 1/\sqrt{L_R C_R}$, and 
the parameters are set as $\zeta = C_L/C_R=0.4$ and $\eta = L_L/L_R=0.9$ (or $\zeta = 0.9$ and $\eta = 0.4$).
(c) Symmetric and (d) antisymmetric eigenmodes at $k_x^\mathrm{(Bloch)} = 0$.
}
 \end{figure}
Figure~\ref{fig:CRLH}(a) shows the unit cell of a CRLH transmission line
with inductors and capacitors.
The unit cell is periodically arranged in the $x$ direction with period $\Delta x$.
When the left-handed parameters are set as
$C_L \rightarrow \infty$ and $L_L \rightarrow \infty$,
the model becomes a conventional transmission line.
The right-handed components are characterized by 
$\omega_0 = 1/\sqrt{L_R C_R}$ and $R_0 = \sqrt{L_R/C_R}$.
For the left-handed components,
we introduce dimensionless parameters $\zeta = C_L/C_R$ and $\eta= L_L/L_R$.
The dispersion relation of a CRLH transmission line for $\zeta=0.4$ and $\eta=0.9$ is depicted in 
Fig.~\ref{fig:CRLH}(b) (see Appendix~\ref{sec:ladder_analysis} for calculation details).
Here, $k_x^\mathrm{(Bloch)}$ and $\omega$ represent the Bloch wavenumber along $x$
and the angular frequency, respectively.
The left-handed components $C_L$ and $L_L$ produce 
the remarkable first band with a negative group velocity, whereas the second band corresponds to the first band for the conventional transmission line with $C_L \rightarrow \infty$ and $L_L \rightarrow \infty$.
From the perspective of electromagnetism,
the negative group velocity indicates 
a left-handed triad $(\tilde{\mathbf{E}},\tilde{\mathbf{H}},\mathbf{k})$
of an electric field, magnetic field, and wave vector, 
whereas the conventional transmission line has a right-handed one.
The name “CRLH transmission line” originates from the fact that it contains both right- and left-handed elements.

\paragraph{Series and Shunt Resonances.}
The CRLH transmission line has mirror symmetry
when the order of the inductor and capacitor positions in 
the series impedance $Z=\jj \omega L_R + 1/(\jj \omega C_L)$ and the shunt admittance $Y=\jj\omega C_R + 1/(\jj \omega L_L)$ are ignored
\footnote{ 
In particular, a CRLH transmission line has mirror symmetry with respect to $x=x_i$ or $x=x_{i+1/2}$, as
described in the effective model of Fig.~\ref{fig:transmission_line}.}.
Because the mirror symmetry still holds at the band edge of the wavenumber space,
the band-edge eigenmodes are classified as symmetric and antisymmetric modes.
We focus on eigenmodes with $k_x^\mathrm{(Bloch)}=0$.
Thus, a symmetric mode must not accompany a series current,
whereas an antisymmetric one leads to a node voltage of zero.
Therefore, the symmetry demands the resonance conditions ($Y=0$ or $Z=0$).
The corresponding resonant modes are depicted
in Figs.~\ref{fig:CRLH}(c) and (d), 
and their angular eigenfrequencies are expressed as follows:
 $\omega_\mathrm{sh}=\omega_0/\sqrt{\eta}$
and $\omega_\mathrm{se}=\omega_0/\sqrt{\zeta}$, respectively.
Clearly, the node potential in Fig.~\ref{fig:CRLH}(c) is symmetric and
the series current in Fig.~\ref{fig:CRLH}(d) is antisymmetric with respect to the mirror reflection.
The band gap appears between $\omega_\mathrm{sh}$ and $\omega_\mathrm{se}$.

\paragraph{Topological Phases.}
\begin{figure}[!t]
 \centering
  \includegraphics{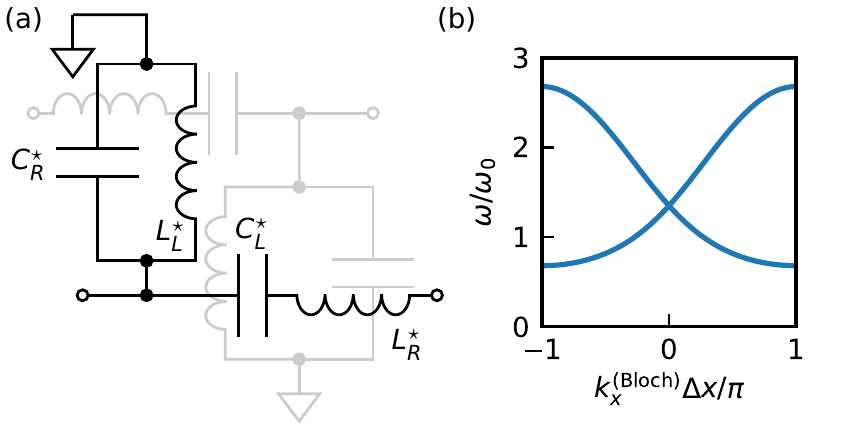}
  \caption{
\label{fig:CRLH_self_dual}
Duality in the CRLH transmission line:
(a) Dual circuit for a CRLH transmission line and (b) dispersion relation of the self-dual CRLH transmission line with $\zeta = \eta = 0.55$.}
 \end{figure}
CRLH transmission lines can be classified into two topological phases:
(i) $\zeta < \eta$ and (ii) $\eta < \zeta$.
When we gradually change the parameter from (i) to (ii),
band inversion occurs at $k_x^\mathrm{(Bloch)}=0$.
This transition induces the flipping of the eigenmode parity at $k^\mathrm{(Bloch)}_x=0$.
Therefore, the parity is a topological integer characterizing the transition.

\paragraph{Duality.}
The two topological phases are implicitly related through the circuit duality.
Consider a dual circuit for the CRLH transmission line 
with respect to a reference 
resistance $R$, as shown in Fig.~\ref{fig:CRLH_self_dual}(a).
We obtain the dual quantities as follows:
\begin{align}
 C_R^\star = \frac{L_R}{R^2}, &\quad L_R^\star=C_RR^2,  \label{eq:45}\\
 C_L^\star = \frac{L_L}{R^2}, &\quad L_L^\star=C_LR^2.  \label{eq:46}
\end{align}
Equation~(\ref{eq:45}) becomes self-dual, 
i.e.,\ $C_R^\star=C_R$ and $L_R^\star=L_R$, 
provided that 
we set $R$ as follows:
\begin{equation}
R= R_0 = \sqrt{\frac{L_R}{C_R}}.  \label{eq:47}
\end{equation}
Under duality transformation, 
the left-handed component is transformed into
\begin{equation}
 C_L^\star = \eta C_R, \quad L_L^\star=\zeta L_R.  \label{eq:48}
\end{equation}
Therefore, the duality transformation induces 
the interchange between $\zeta$ and $\eta$.
In particular, the swap maintains the shape of the dispersion relation,
whereas the symmetry of eigenmodes is interchanged at $k_x^\mathrm{(Bloch)}=0$.
The self-duality characterizes the transition point of the two phases
as $\zeta=\eta$. Figure~\ref{fig:CRLH_self_dual}(b) shows the dispersion curve for $\zeta = \eta = 0.55$.
We can clearly observe the Dirac-point formation at $k_x^\mathrm{(Bloch)}=0$,
which is protected by self-duality.

\paragraph{Bulk--Reactance Correspondence.}
Each topological phase has
a definite sign of the Bloch reactance inside the band gap.
We prove this bulk--reactance correspondence from a circuit-theoretical perspective.
The symmetries of the eigenmodes demand
$Z^\mathrm{(Bloch)}(\omega_\mathrm{sh})=\infty$
and $Z^\mathrm{(Bloch)}(\omega_\mathrm{se})=0$, as shown in Figs.~\ref{fig:CRLH}(c) and (d), respectively. 
The band gap is denoted by $(\omega_1, \omega_2)$ with 
$\omega_1 = \operatorname{min}(\omega_\mathrm{sh}, \omega_\mathrm{se})$
and $\omega_2 = \operatorname{max}(\omega_\mathrm{sh}, \omega_\mathrm{se})$.
In $\omega_1 < \omega < \omega_2$, 
$Z^\mathrm{(Bloch)}$ is purely imaginary, 
and $Z^\mathrm{(Bloch)}$ has no zeros or poles.
Because $\operatorname{Im}[Z^\mathrm{(Bloch)}]$ monotonically increases with an increase in $\omega$ from the reactance theorem,
the band-gap behavior in $\omega_1<\omega<\omega_2$ is determined as follows:
(i) capacitive response $\operatorname{Im}[Z^\mathrm{(Bloch)}(\omega)]<0$ 
for $\omega_\mathrm{sh}< \omega_\mathrm{se}$
or (ii) inductive response $\operatorname{Im}[Z^\mathrm{(Bloch)}(\omega)]>0$
for $\omega_\mathrm{sh}> \omega_\mathrm{se}$.
Therefore, we have completed the proof.
As shown in Appendix~\ref{sec:sym_constraint}, 
the definite sign of the Bloch reactance inside a band gap
can be established even in a continuous (distributed-circuit) model.

Note that the parity at $k_x^\mathrm{(Bloch)}\Delta x = \pm \pi$ is kept unchanged under the transition
between the two phases of (i) and (ii). 
Thus, it does not affect the surface impedance inside the band gap.
The parity at $k_x^\mathrm{(Bloch)}\Delta x = \pm \pi$ depends on the choice of the unit cell.
In fact, $\Pi$ and T units give different parities at $k_x^\mathrm{(Bloch)}\Delta x = \pm \pi$ (see Appendix~\ref{sec:ladder_analysis}).
Therefore, the conventional formula involving the Zak phase \cite{Xiao2014} 
cannot be naively applied to plasmonic systems.

\subsection{CRLH Model for Dielectric--Metal Transition \label{sec:CRLH_dielectric_metal}}

\begin{figure*}
 \centering
  \includegraphics{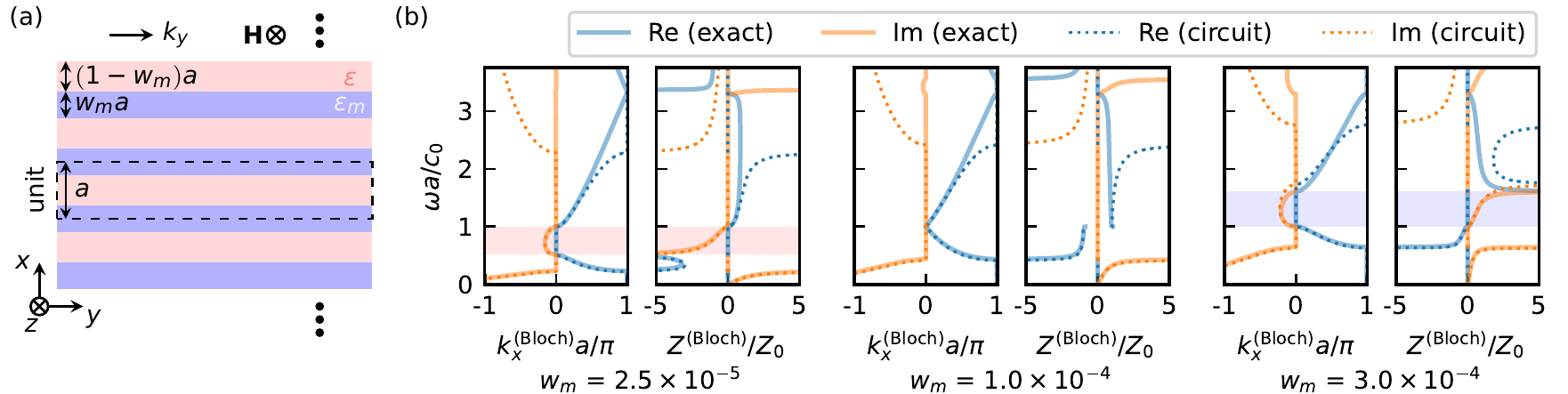}
  \caption{
\label{fig:dielectric_metal_transition}
Continuous transition between dielectric and metal:
(a) Configuration of a plasmonic/photonic crystal
with dielectric permittivity $\varepsilon$ and Drude permittivity $\varepsilon_m = \varepsilon_0 [1- (\omega_p/\omega)^2]$.
The dashed line denotes a unit cell with period $a$.
(b) TM Bloch wavenumber and impedance with respect to the frequency 
for different metal portions ($w_m$).
The solid and dotted lines indicate
the exact and circuit-model calculations, respectively.
The band gap between the first and second bands is colored
(light pink: capacitive; light blue: inductive).
The calculation parameters were set as
$\varepsilon = \varepsilon_0$, $k_y a = 1$, and $\omega_p a/c_0 = 100$.
The Bloch impedance 
normalized by $Z_0=\sqrt{\mu_0/\varepsilon_0}$
was evaluated at the bottom of the unit cell.
}
 \end{figure*}

We show that the CRLH transmission line reflects underlying physics of the topological dielectric--metal transition.
Consider dielectric and metallic layers, as indicated by the dashed box in Fig.~\ref{fig:dielectric_metal_transition}(a).
The bottom of the unit cell is selected as the center of the metal
to simplify the Bloch impedance.
We analyze a TM wave with wavenumber $k_y>0$ in the $y$ direction.
The period of the unit cell in the $x$ direction is denoted as $a$.
The dielectric layer has thickness $(1-w_m)a$ and permittivity $\varepsilon$.
The speed of light in the dielectric layer is given as follows: $c=1/\sqrt{\varepsilon \mu_0}$.
The permittivity of the metallic layer with thickness $w_m a$ is given by the Drude model with Eq.~(\ref{eq:42}).
By changing the metal portion $w_m$,
we can induce a topological phase transition \cite{Yang2020}.
To derive a CRLH model theoretically, 
we focus on $\omega \ll \omega_p$, in contrast to Ref.~\onlinecite{Yang2020}.
To induce band inversion in $\omega \ll \omega_p$, 
$c k_y \ll \omega_p$ is required.

The TM mode propagation along $x$ 
in the dielectric and metal can be modeled 
using one-dimensional circuits, as 
described in Appendices~\ref{sec:analogy_circuit_EM} and \ref{sec:dielectric_metal_models}.
In developing the circuit model of the binary unit cell, 
we assume that the positions 
of the shunt admittance and series impedance can be exchanged freely in the unit cell.
As shown in Appendix~\ref{sec:YZswap}, this assumption is justified when the shunt current is sufficiently small compared with the series current.
Then, we obtain a CRLH model for the unit cell as shown in Fig.~\ref{fig:dielectric_metal_transition}(a).
The circuit parameters are given as follows:
\begin{align}
 L_R&= \mu_0 a + \frac{{k_y}^2 w_m a}{\varepsilon_0{\omega_p}^2},  \label{eq:49}\\
 C_R &=\left[\varepsilon(1-w_m)+\varepsilon_0 w_m \right]a,  \label{eq:50}\\
 L_L &=  \frac{1}{{\omega_p}^2\varepsilon_0 w_m a},  \label{eq:51}\\
 C_L &= \frac{1}{(c k_y)^2 \mu_0 (1-w_m) a}.  \label{eq:52}
\end{align}

Next, we analyze the continuous and simplified CRLH models 
and compare their results. 
First, we describe our calculation setup.
We select $\varepsilon = \varepsilon_0$
and set $k_y a = 1$ and $\omega_p a/c_0 = 100$.
For a given frequency, the Bloch wavenumber $k^\mathrm{(Bloch)}_x$ along $x$
is calculated. 
The Bloch impedance $Z^\mathrm{(Bloch)}$ in the exact model is 
evaluated at the bottom of the unit cell (i.e.,\ the center of the metal).
The continuous model is treated as described in Appendix~\ref{sec:periodic}, 
whereas the circuit model is analyzed based on Appendix~\ref{sec:ladder_analysis}.
The circuit model exhibits ambiguity in defining $Z^\mathrm{(Bloch)}$.
It is calculated for the $\Pi$ unit so that $Z^\mathrm{(Bloch)}$ is inductive
near $\omega=0$ (see Appendix~\ref{sec:ladder_analysis}).
In both systems, we select the solution that does not diverge at $x=+\infty$.
Then, the following condition is imposed:
\begin{equation}
 \left|\exp\left(-\jj k_x^\mathrm{(Bloch)} a\right)\right| \leq 1.  \label{eq:53}
\end{equation}
To simplify the plots, we only include the modes obeying the following condition:
\begin{equation}
\operatorname{Im}\left[\exp\left(-\jj k_x^\mathrm{(Bloch)} a\right)\right]\leq 0.  \label{eq:54}
\end{equation}
Because the CRLH transmission line has 
time-reversal symmetry,
we can produce solutions satisfying 
$\operatorname{Im}[\exp(-\jj k_x^\mathrm{(Bloch)} a)] > 0$
by applying the time-reversal operation 
to the solutions satisfying Eq.~(\ref{eq:54}).
The symmetry consideration indicates the following properties:
(i) the Bloch wavenumber and impedance are real or purely imaginary
and (ii) the Bloch impedance must be 0 or $\pm \infty$ at the band edge $k_x^\mathrm{(Bloch)}=0,\ \pm \pi/a$, as described in Appendix~\ref{sec:sym_constraint}.
Because $k^\mathrm{(Bloch)}_x$ and $Z^\mathrm{(Bloch)}$ can be imaginary, 
we plot their real and imaginary parts with respect to the frequency.

Figure~\ref{fig:dielectric_metal_transition}(b) 
shows the Bloch wavenumber and impedance evaluated using the two methods.
The circuit model solution (dotted) agrees well with 
the exact solution (solid) below the bottom part of the second real band.
Therefore, the circuit model substantially captured
the band inversion between the first and second bands.
We stress that the circuit model does \textit{not} involve any fitting parameters.
At the high frequencies, the circuit model does not 
approximate the exact solution well.
This disagreement is reasonable considering that the circuit model involves only 
a few degrees of freedom, whereas the exact model has an infinite degree of freedom.
In fact, the exact treatment produces an infinite number of bands, 
whereas the circuit model only yields two bands.
The shunt--series swapping to obtain the CRLH model is justified as follows.
In the circuit model, the series impedances of the dielectric and metallic regions are represented by $Z_D$ and $Z_M$, respectively,
and the shunt admittances are represented by $Y_D$ and $Y_M$, respectively.
Because we focus on $w_m\ll 1$, $|Z_M|$ becomes small.
Then, $|{Z_M}^{-1}+{Z_D}^{-1}|$ becomes large 
except for $\omega = \omega_\mathrm{se} \approx c_0 k_y$
at which ${Z_D}$ exhibits the resonance.
Therefore, $|Y_D| \ll |{Z_M}^{-1}+{Z_D}^{-1}|$ 
is expected at $\omega \ne \omega_\mathrm{se}$.
At $\omega = c_0 k_y$, $|Z_D|$ becomes zero, and 
$|Z_D| \ll |{Y_D}^{-1} + {Y_M}^{-1}|$ holds.
In conclusion, both cases satisfy the swapping condition presented in Appendix~\ref{sec:YZswap}.

We explain the transition in terms of the CRLH model.
Using Eqs.~(\ref{eq:49})--(\ref{eq:52}) with $\varepsilon=\varepsilon_0$,
the CRLH parameters are determined as follows:
\begin{align}
 \omega_0 &= \frac{c_0}{a\sqrt{1 + w_m\left(\frac{c_0 k_y}{{\omega_p}}\right)^2}}, \label{eq:55}\\ 
R_0 &= Z_0 \sqrt{1 + w_m\left(\frac{c_0 k_y}{{\omega_p}}\right)^2},  \label{eq:56}\\
 \zeta& = \frac{1}{{k_y}^2a^2 (1-w_m)}, \label{eq:57}\\
 \eta&= \frac{{c_0}^2}{{\omega_p}^2 a^2 w_m \left[1 + w_m\left(\frac{c_0 k_y}{{\omega_p}}\right)^2\right]}.\label{eq:58}
\end{align}
Therefore, $\omega_\mathrm{sh}<\omega_\mathrm{se}$ ($\zeta < \eta$) holds for small $w_m$,
whereas $\omega_\mathrm{sh}>\omega_\mathrm{se}$ ($\zeta > \eta$) holds for large $w_m$.
Figure~\ref{fig:dielectric_metal_transition}(b) presents 
the crossover between them.
The mirror symmetry leads to
$Z^\mathrm{(Bloch)} = \infty$ at $\omega_\mathrm{sh}$,  
whereas the antisymmetry results in 
$Z^\mathrm{(Bloch)}=0$ at $\omega_\mathrm{se}$.
If $\omega_\mathrm{sh} < \omega_\mathrm{se}$ is satisfied,
$\operatorname{Im}[Z^\mathrm{(Bloch)}]|_{\omega=\omega_\mathrm{sh}+0^+} = -\infty$ 
is required for the reactance theorem.
Hence, the gap behaves capacitively; 
$\operatorname{Im}[Z^\mathrm{(Bloch)}]<0 $ 
for $\omega_\mathrm{sh} < \omega < \omega_\mathrm{se}$.
By contrast,
$\omega_\mathrm{se} < \omega_\mathrm{sh}$ demands
$\operatorname{Im}[Z^\mathrm{(Bloch)}]|_{\omega=\omega_\mathrm{sh}-0^+} = +\infty$;
thus, the gap behaves inductively: 
$\operatorname{Im}[Z^\mathrm{(Bloch)}]> 0 $ 
for $\omega_\mathrm{se} < \omega <\omega_\mathrm{sh}$.
These bulk--reactance correspondences are confirmed for the filled regions in Fig.~\ref{fig:dielectric_metal_transition}(b). 

Now, we consider $w_m\rightarrow 0^+$ and $w_m\rightarrow 1^-=1-0^+$.
The dielectric limit $w_m\rightarrow 0^+$ gives 
$\eta\rightarrow \infty$, resulting in
$\omega_\mathrm{sh}\rightarrow 0^+$ and $\omega_\mathrm{se}\rightarrow c_0 k_y$, which is the cutoff frequency of the dielectric material.
The metallic limit $w_m\rightarrow 1^-$
leads to 
$\zeta\rightarrow \infty$,
resulting in $\omega_\mathrm{se}\rightarrow 0$
and $\omega_\mathrm{sh}\rightarrow \omega_p$.
Therefore, the plasmonic gap forms in $0< \omega <\omega_p$.
We need not consider 
the approximation condition of $\omega\ll \omega_p$ at $w_m\rightarrow 1^-$,
because the additional capacitor ($C_\mathrm{se}'$ discussed in Appendix~\ref{sec:dielectric_metal_models}) does not contribute to the resonance frequencies.
Shunt and series zero modes with different symmetries at $w_m=0^+$ and $1^-$
are responsible for the frequency responses in the quasistatic regime,
which are given by $\operatorname{Im}Z^\mathrm{(Bloch)}|_{\omega= 0^+} = -\infty$ 
and $\operatorname{Im}Z^\mathrm{(Bloch)}|_{\omega= 0} = 0$, respectively.

Now, we clarify the difference between the bulk--edge correspondence 
established in Sec.~\ref{sec:bulk-edge}
and the bulk--reactance correspondence proven here.
The bulk--reactance correspondence indicates 
capacitive and inductive behaviors for $\omega\rightarrow 0^+$ 
for $w_m\rightarrow 0^+$ and $1^-$, respectively.
Therefore, it is implicitly related to an LC resonance; 
however, it does not generally ensure the existence of such a resonance.
The existence of surface plasmon polaritons on metals
is explained by the previous bulk--edge correspondence.

\subsection{Electric and Magnetic Zero Modes \label{sec:em_zero_modes}}

In this subsection, we identify localized zero modes, 
which produce the flat zero bands at the dielectric and metallic limits. 
They are responsible for the zero resonances discussed
in Sec.~\ref{sec:bulk-edge_from_circuit}
and give the physical origin of the extraordinary first band with a negative group velocity.

\begin{figure}[bt]
 \includegraphics{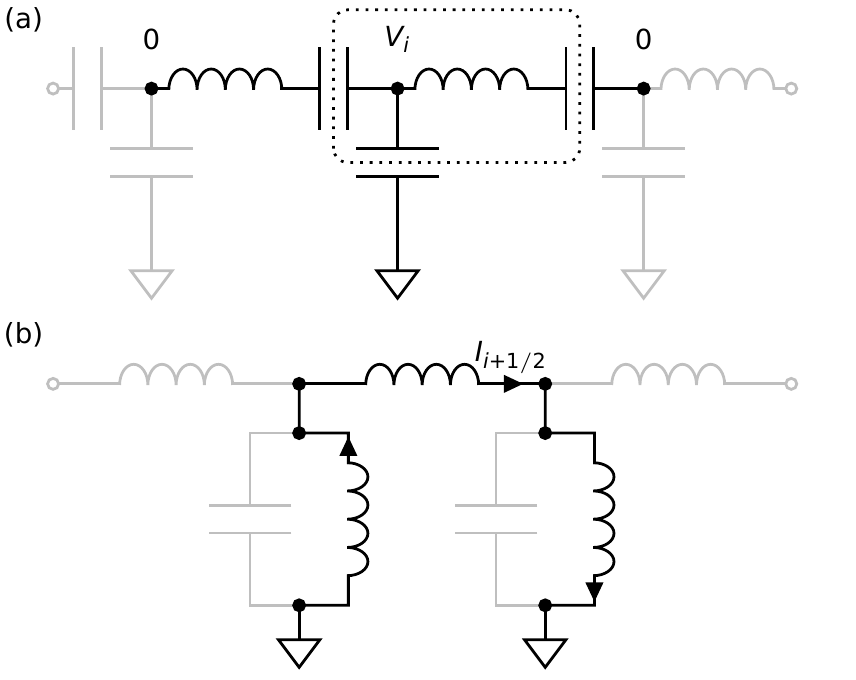}
  \caption{Zero modes in the CRLH circuit model:
(a) capacitive and (b) inductive zero modes for $\eta\rightarrow \infty$ and $\zeta\rightarrow \infty$, respectively. In (a), the node voltage satisfies $V_i\ne 0$ and $V_l = 0 $ ($l\ne i$), where the corresponding cut set across only capacitors 
is depicted as a dotted line. In (b), a current $I_{i+1/2}$ flows along a loop comprising inductors only.
\label{fig:circuit_zero_modes}}
\end{figure}

From the perspective of the circuit model, 
responses at $w_m = 0$ and $1$ originate from the zero modes shown in Fig.~\ref{fig:circuit_zero_modes}. 
These zero modes are bulkily degenerated and form flat zero bands.
They are dual with each other and have constraints on 
the total charge in a cut set comprising capacitors
or the flux penetrating a loop comprising inductors.
These constraints can be interpreted as
DC freezing, under the limit of $\eta \rightarrow \infty$ or $\zeta\rightarrow \infty$.
Next, we construct the corresponding states in the continuous model.

\begin{figure}[bt]
 \includegraphics{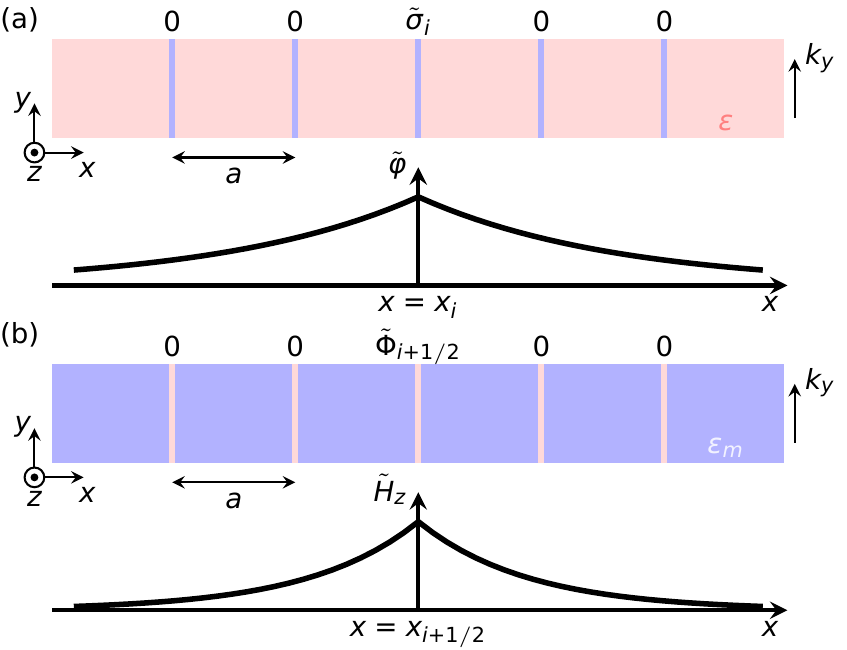}
  \caption{Zero modes in the continuous model:
(a) Electric and (b) magnetic zero modes localized 
at $x=x_i=i a$ and $x=x_{i+1/2} = (i+1/2)a $
for $w_m=0^+$ and $1^-$, respectively. 
The electric charge (surface density)
at $x_i$ is denoted as $\tilde{\sigma}_i\exp(-\jj k_y y)$
for $i\in \mathbb{Z}$, and 
the magnetic flux (line density) near $x_i = (i+1/2)a$ is represented by
$\tilde{\Phi}_{i+1/2}\exp(-\jj k_y y)$.
In (a), the electrostatic potential $\tilde{\varphi}(x) \exp(-\jj k_y y)$
with $\tilde{\varphi}(x)\propto \exp\big[-k_y |x-x_i|\big]$
is induced by the charge layer satisfying $\tilde{\sigma}_i\ne 0$ 
and $\tilde{\sigma}_l = 0 $ ($l\ne i$).
In (b), the magnetic field $\tilde{H}_z(x) \exp(-\jj k_y y)$
with $\tilde{H}_z(x)\propto \exp\left[-\sqrt{{k_y}^2 +(\omega_p/c_0)^2} |x-x_{i+1/2}|\right]$ is produced by the trapped magnetic flux near $x_{i+1/2}$.
\label{fig:continuous_zero_modes}}
\end{figure}

First, we investigate $w_m\rightarrow 0^+$.
We consider the F$_0$ matrix of a single metallic layer 
at $x=0$ with an infinitely thin thickness $d\rightarrow 0^+$.
From Eq.~(\ref{eq:35}) with $\varepsilon_m \rightarrow -\infty$ for $\omega\rightarrow 0^+$, the
$E_y$ continuity is deduced as $E_y|_{x=0^-} = E_y|_{x=0^+}$.
Conversely, $D_x$ may have discontinuity at $x=0$, 
which represents the charge degree of freedom at the layer.
Although the charge cannot exist for $d=0$, 
insertion of an infinitely thin metallic layer at $d=0^+$ adds the degree of freedom.
The infinitely thin layer works only for $\omega=0$ and does not contribute to the frequency response for $\omega >0$,
considering that the permittivity is finite at $\omega >0$.
Therefore, the insertion is interpreted as a \textit{zero-mode addition}.
For a periodic system with $w_m=0^+$, we can construct a zero mode 
generated by a charge located only on a single layer
by using the solution of Eq.~(\ref{eq:23}).
The constructed mode is shown in Fig.~\ref{fig:continuous_zero_modes}(a)
and corresponds to Fig.~\ref{fig:circuit_zero_modes}(a).
The zero modes compose the flat zero band considering that they form at all layers.

Second, we analyze $w_m=1^-$.
We start with $w_m=1$.
From Eqs.~(\ref{eq:61})--(\ref{eq:63}), the
$\tilde{H}_z$ inside the Drude metal obeys
\begin{equation}
 \odv{^2\tilde{H}_z}{x^2} = ({k_y}^2 - \omega^2 \varepsilon_m \mu_0)\tilde{H}_z,  \label{eq:59}
\end{equation}
which is reduced to $\dd^2\tilde{H}_z/\dd x^2 = [{k_y}^2 + (\omega_p/c_0)^2] \tilde{H}_z$ for $\omega\rightarrow 0^+$.
Therefore, the zero-frequency solutions are given by $\exp\left(\pm \sqrt{{k_y}^2+(\omega_p/c_0)^2}\, x\right)$.
By contrast, electric fields inside the metal 
approach zero for $\omega\rightarrow 0^+$ from Eqs.~(\ref{eq:61}) and (\ref{eq:62}).
Now, we place a single gap at $x=0$ inside the Drude metal. 
It has an infinitely thin thickness $d=0^+$
and permittivity $\varepsilon>0$. 
From Eq.~(\ref{eq:69}), we conclude that $\tilde{H}_z$ is continuous at $x=0$ for $\omega\rightarrow 0^+$.
However, $\tilde{E}_y$ may possess a discontinuity at $x=0$,
which indicates a discontinuous $\dd \tilde{H}_z/\dd x$ at $x=0$ 
from Eq.~(\ref{eq:62}).
The discontinuity is interpreted as the magnetic-flux degree of freedom.
Therefore, the insertion of an infinitely thin dielectric slab into a metal involves a zero-mode addition,
whereas the frequency response is kept unchanged for $\omega>0$.
Finally, we can construct a magnetic zero mode, as shown
in Fig.~\ref{fig:continuous_zero_modes}(b), corresponding to Fig.~\ref{fig:circuit_zero_modes}(b).
Each layer has an individual zero mode, resulting in 
the flat zero band.

In summary, electric and magnetic zero modes exist at $w_m=0^+$ and $1^-$ and are responsible for the zero resonances discussed in Sec.~\ref{sec:bulk-edge_from_circuit}. They
highlight the essential difference between dielectrics and metals.
The change in $w_m$ from $0^+$ to $1^-$ induces the interchange 
between the electric and magnetic zero modes, resulting in the phase transition.

\section{\label{sec:conclusion} Conclusions}

In this study, we revealed the origin of surface Maxwell waves.
The results indicated that the surface plasmon polaritons originate from robust electric zero modes with $\mathcal{CM}_x$ symmetry.
The similar zero modes can be experimentally excited at a temporal boundary.
Owing to the Keller--Dykhne self-duality, 
the zero modes possess real-space polarization rotation, 
which is related to the surface impedance.
Furthermore, the bulk--edge correspondence can be summarized as follows:
a surface mode exists between the inductive and capacitive surfaces.
However, this statement cannot be proved when focusing on electric zero modes alone.
To solve this problem, we introduced magnetic response and frequency dispersion.
Then, we generally proved the bulk--edge correspondence
with circuit-theoretical considerations.
Finally, we proposed a CRLH transmission line with duality,
fully explaining the essential physics underlying the topological transition between a metal and a dielectric material.
In summary, surface plasmon polaritons are comprehensively understood from two different viewpoints:
symmetry protection and bulk--edge correspondence, which are related to each other.
The elucidated physics of surface Maxwell waves 
is now within the reach of experimental verification.

\begin{acknowledgments}
The authors thank A.~Okamoto, J.~Matsudaira, Y.~Shikano, H.~Maruhashi, and A.~Sanada for their fruitful discussions. This work was supported by 
JST, PRESTO (Grant No. JPMJPR20L6), and JSPS KAKENHI 
(Grant Nos. 19K05304, 20K14374, 20H01845, 22K04964, and 22H00108).
\end{acknowledgments}

\nocite{*}

\appendix

\section{Basic Equations of Surface Plasmon Polaritons \label{sec:basicEqs}}

Assume that the electric permittivity $\varepsilon(x)$ and 
magnetic permeability $\mu(x)$ depend on $x$ alone.
Consider a TM wave with wavenumber $k_y$ in the $y$ direction.
It has a $z$-component magnetic field 
$H_z(t, x,y) = \tilde{H}_z(x)\exp[\jj (\omega t -k_y y)] + \mathrm{c.c.}$
with the complex amplitude $\tilde{H}_z(x)$.
Similarly, we define the complex amplitudes 
of the electric displacement $\tilde{\mathbf{D}}$, electric field $\tilde{\mathbf{E}}$,
and magnetic flux density $\tilde{\mathbf{B}}$.
Here, we use the convention that 
a variable with a tilde dependent only on $x$ 
always represents the complex amplitude omitting $\exp(-\jj k_y y)$.
The laws of Gauss, Amp\'{e}re--Maxwell, and Faraday are given by the following equations:
\begin{align}
 \odv{\tilde{D}_x}{x} - \jj k_y \tilde{D}_y &= 0,  \label{eq:60}\\
 -\jj k_y \tilde{H}_z  &= \jj \omega \tilde{D}_x,  \label{eq:61}\\
 -\odv{\tilde{H}_z}{x}  &= \jj \omega \tilde{D}_y,  \label{eq:62}\\
 \odv{\tilde{E}_y}{x} +\jj k_y \tilde{E}_x &= -\jj \omega \tilde{B}_z.  \label{eq:63}
\end{align}
Note that these equations are not independent.
Eq.~(\ref{eq:61}) and Eq.~(\ref{eq:62}) lead to Eq.~(\ref{eq:60}) for $\omega\ne 0$, although Eq.~(\ref{eq:60}) should be independently demanded at $\omega=0$.

\section{F Matrix of Slab for TM Waves \label{sec:F-matrix}}

A uniform slab is located in $x\in [0,d]$ ($d>0$) 
with the real scalar permittivity $\varepsilon$ and permeability $\mu$.
We derive the relationship between the fields at $x=0$ and $x=d$.
Consider a TM wave, where the magnetic field is oriented in the $z$ direction.
The angular frequency and wavenumber along the $y$-axis are represented by
$\omega$ and $k_y$, respectively.
The complex amplitude of the $z$ component of the magnetic field is represented as
\begin{equation}
 \tilde{H}_z(x) = \tilde{H}_1 \exp(-\jj k_x x) +\tilde{H}_2 \exp(\jj k_x x),  \label{eq:64}
\end{equation}
where we omit the temporal variation $\exp(\jj\omega t)$ and spatial variation $\exp(-\jj k_y y)$.
The wavenumber $k_x$ satisfies
\begin{equation}
{k_x}^2 + {k_y}^2 = \varepsilon\mu \omega^2.  \label{eq:65}
\end{equation}
Using the Amp\'{e}re--Maxwell equation, we obtain the $y$ component of the electric-field amplitude as
\begin{align}
 \tilde{E}_y(x) &= -\frac{1}{\jj\omega\varepsilon} \odv{\tilde{H}_z}{x}\nonumber \\
&= 
Z \left[\tilde{H}_1 \exp(-\jj k_x x) -\tilde{H}_2 \exp(\jj k_x x) \right],  \label{eq:66}
\end{align}
where we define the TM wave impedance $Z$ as
\begin{equation}
 Z= \frac{k_x}{\omega \varepsilon}.  \label{eq:67}
\end{equation}
From Eqs.~(\ref{eq:64}) and (\ref{eq:66}), the fields at $x=0$ and $x=d$ are related as follows:
\begin{equation}
\begin{bmatrix}
\tilde{E}_y(0) \\
\tilde{H}_z(0)
\end{bmatrix} 
=
F
\begin{bmatrix}
\tilde{E}_y(d) \\
\tilde{H}_z(d)
\end{bmatrix}.  \label{eq:68}
\end{equation}
The F matrix of the slab with thickness $d$ is given as
\begin{equation}
F = 
\begin{bmatrix}
 \cosh \theta & Z\sinh \theta\\
Z^{-1} \sinh\theta & \cosh\theta
\end{bmatrix},  \label{eq:69}
\end{equation}
where $\theta =\jj k_x d$ is the propagation constant.
There are two sign choices for $k_x$ in Eq.~(\ref{eq:65});
however, both give the same F matrix. 
Conventionally, $k_x$ is defined so that the $\exp(-\jj k_x x)$ term represents the physical mode in a half-space, as discussed in Appendix~\ref{sec:uniform_half}.
Equation~(\ref{eq:69}) satisfies
\begin{equation}
 \det F = 1,  \label{eq:70}
\end{equation}
which is related to the reciprocity \cite{Collin1992}.

\section{Effective Response of Uniform Half-Space \label{sec:uniform_half}}

We characterize the effective response on $x=0$ 
for a uniform half-space with real scalar permittivity $\varepsilon$ and permeability  $\mu$
located in $x\geq 0^+$.
We use the convention that the $\exp(-\jj k_x x)$ terms become physical in the half-space.
We separately consider two cases: a real $k_x$ and a purely imaginary $k_x$.
For a real $k_x$, there are two propagating modes in the $x$ direction.
One carries energy from $x=0$ to infinity, whereas the other brings energy from $x=+\infty$ to $x=0$.
When we treat the effective response on $x=0$ for the half-space, 
the latter should be ignored.
Therefore, $\operatorname{Re} Z\geq 0$ is needed
to ensure that $\exp(-\jj k_x x)$ 
represents the energy flow from $x=0$ to $x=+\infty$.
Next, we treat the case of the imaginary $k_x$.
When the $\exp(-\jj k_x x)$ terms should be finite at $x\rightarrow +\infty$, 
$\operatorname{Im}k_x \leq 0$ is required.
In summary, the wavenumber along the $x$-axis is defined as follows:
\begin{equation}
 k_x = 
\begin{cases}
\sqrt{\varepsilon\mu \omega^2- {k_y}^2} & (\varepsilon> 0,\ \mu>0,\ \varepsilon\mu \omega^2 > {k_y}^2)\\
-\sqrt{\varepsilon\mu \omega^2- {k_y}^2} & (\varepsilon<0,\ \mu<0,\ \varepsilon\mu \omega^2 > {k_y}^2)\\
-\jj\sqrt{{k_y}^2 - \varepsilon\mu\omega^2} & (\varepsilon\mu \omega^2\leq  {k_y}^2)
\end{cases}  \label{eq:71}
\end{equation}
According to Eq.~(\ref{eq:71}),
the effective response of the half-space is represented by $Z$, which is defined in Eq.~(\ref{eq:67}).

\section{Bloch Analysis of Periodic Binary Dielectrics \label{sec:periodic}}

With regard to Secs.~\ref{sec:robustness} and \ref{sec:CRLH_dielectric_metal},
we consider a periodic arrangement of binary slabs composed of $A$ and $B$.
Let $A$ and $B$ be two dielectric materials with permittivities $\varepsilon_A$ and $\varepsilon_B$, respectively.
The unit cell of thickness $a=d_A+d_B$ is filled with $A$ and $B$ in $0\leq x\leq d_A$ and $d_A\leq x \leq d_A+d_B$, respectively. We periodically arrange the unit cell in $x\geq 0$.
The wavenumber in $X=A, B$ is 
denoted as $k_x^{(X)}$.
We define $\varphi_A = k_x^{(A)} d_A$
and $\varphi_B = k_x^{(B)} d_B$.
The unit F matrix obtained from the multiplication is given as
\begin{multline}
 F^\mathrm{(unit)}  =
\begin{bmatrix}
F_{11}&F_{12}\\
F_{21}&F_{22}
\end{bmatrix}\\
=
\begin{bmatrix}
\cos \varphi_A & \jj Z_A \sin \varphi_A\\
\jj Y_A \sin \varphi_A & \cos \varphi_A
\end{bmatrix}
\begin{bmatrix}
\cos \varphi_B & \jj Z_B \sin \varphi_B\\
\jj Y_B \sin \varphi_B & \cos \varphi_B
\end{bmatrix}.  \label{eq:72}
\end{multline}
Here, we introduce the wave impedance as
\begin{equation}
 Z_X=\left(Y_X\right)^{-1} = \frac{k_x^{(X)}}{\omega \varepsilon_X}.  \label{eq:73}
\end{equation}
The Bloch wavenumber $k^\mathrm{(Blcoh)}_x$ is given as
\begin{equation}
 F^\mathrm{(unit)}
\begin{bmatrix}
 \tilde{E}\\ \tilde{H}
\end{bmatrix} 
=\exp\left[\jj k^\mathrm{(Blcoh)}_x a\right]
\begin{bmatrix}
 \tilde{E}\\ \tilde{H}
\end{bmatrix}.  \label{eq:74}
\end{equation}
When we focus on the decaying wave in the positive $x$ direction, 
\begin{equation}
 \operatorname{Im}\left[k^\mathrm{(Blcoh)}_x\right] < 0  \label{eq:75}
\end{equation}
is required.
Using $\det F^\mathrm{(unit)}=1$,
we obtain
\begin{equation}
 \exp\left[\jj k^\mathrm{(Bloch)}_x a\right]  = \frac{\operatorname{tr} F^\mathrm{(unit)} \pm\sqrt{(\operatorname{tr} F^\mathrm{(unit)})^2-4}}{2} > 1.  \label{eq:76}
\end{equation}
Here, the trace can be calculated as
\begin{multline}
 \operatorname{tr} F^\mathrm{(unit)} = F_{11}+F_{22}\\
=
2\cos \varphi_A \cos \varphi_B -\left(\frac{Z_A}{Z_B}+\frac{Z_B}{Z_A}\right)\sin \varphi_A\sin \varphi_B.  \label{eq:77}
\end{multline}
The Bloch impedance on $x=0$ for $x\geq 0$ is given as
\begin{equation}
Z^\mathrm{(Bloch)} = \frac{\tilde{E}}{\tilde{H}}=\frac{F_{12}}{\exp\left[\jj
  k^\mathrm{(Bloch)}_x a\right] - F_{11}}.  \label{eq:78}
\end{equation}

\section{Effective-Medium Approximation of Periodic Layers \label{sec:effective_media}}

For Secs.~\ref{sec:robustness} and \ref{sec:emprical}, we introduce the effective-medium approximation for periodic layers.
We choose $x_1$, $x_2$, and $\cdots$ satisfying $x_0=0< x_1 < x_2< \cdots < x_n$.
The uniform slabs with permittivity $\varepsilon_i$ 
occupy the region of $x\in [x_{i-1},x_i]$ ($i=1,2,\cdots, n$).
The slab width is denoted as $d_i = x_i - x_{i-1}$.
A unit cell with thickness $a= \sum_{i=1}^n d_i$ consists these slabs.
The unit cells are periodically arranged in $x$.
If the typical length of spatial field variation is significantly longer than $a$,
the system behaves as an anisotropic medium with permittivities $\varepsilon_x$
and $\varepsilon_y$ along the $x$ and $y$-directions, respectively.

First, we derive the effective parameter $\varepsilon_x$.
We apply an electric displacement $D$ along $x$.
From the boundary condition, $D$ is constant along $x$.
Considering $E_\mathrm{eff} a = \sum_i {\varepsilon_i}^{-1} D d_i$
with $E_\mathrm{eff} = {\varepsilon_x}^{-1} D$, 
we obtain
\begin{equation}
 \frac{1}{\varepsilon_x} = \sum_{i=1}^n \frac{1}{\varepsilon_i} \frac{d_i}{a}.  \label{eq:79}
\end{equation}
Thus, we must average ${\varepsilon_i}^{-1}$ to calculate ${\varepsilon_x}^{-1}$.

Second, we determine the effective parameter $\varepsilon_y$ when applying 
an electric field in $y$.
The $y$-component electric field $E$ is constant along $x$ 
owing to the boundary conditions.
From $D_\mathrm{eff} a = \sum_i \varepsilon_i E d_i$
with $D_\mathrm{eff} = \varepsilon_y E$,
we obtain
\begin{equation}
 \varepsilon_y = \sum_{i=1}^{n} \varepsilon_i \frac{d_i}{a},  \label{eq:80}
\end{equation}
which indicates that 
the average of $\varepsilon_i$ gives 
$\varepsilon_y$.

\section{Surface Plasmon Polaritons on Anisotropic Medium \label{sec:anisotropic_spp}}

For Secs.~\ref{sec:robustness} and \ref{sec:emprical},
we derive the dispersion relation of surface plasmon polaritons between anisotropic media
with diagonalized permittivity tensor components along $x$, $y$, and $z$.
The permeability is assumed to be $\mu_0$ in the entire space.
In $x\geq 0^+$, we consider a dielectric material with the permittivity-tensor components
$\varepsilon_x^{(1)}>0$ and $\varepsilon_y^{(1)}>0$ along the $x$- and $y$-axes, respectively.
In $x\leq 0^-$, we place a metallic material with the permittivity-tensor components
$\varepsilon_x^{(2)}<0$ and $\varepsilon_y^{(2)}<0$ along the $x$- and $y$-axes, 
respectively.
We focus on a TM wave with wavenumber $k_y>0$ along $y$.
From Eqs.~(\ref{eq:61})--(\ref{eq:63}), we obtain
\begin{equation}
 \odv{^2\tilde{H}_z}{x^2} = \epsilon_y^{(i)} \left[\frac{1}{\epsilon_x^{(i)}} {k_y}^2 - {k_0}^2 \right] \tilde{H}_z,  \label{eq:81}
\end{equation}
where we use the relative permittivity $\epsilon_a^{(i)} = \varepsilon_a^{(i)}/\varepsilon_0$.
When we assume $\exp\left(-\kappa^{(1)}x\right)$ ($x\geq 0^+$)
and $\exp\left(-\kappa^{(2)}x\right)$ ($x\leq 0^-$), the decay constant 
$\kappa^{(i)}$ can be calculated as
\begin{equation}
\kappa^{(i)} = \sqrt{
\epsilon_y^{(i)} \left[\frac{1}{\epsilon_x^{(i)}} {k_y}^2 - {k_0}^2  \right]}.  \label{eq:82}
\end{equation}
Let $Z_1$ and $Z_2$
be the surface impedances on $x=0$ for $x\geq 0^+$ and $x\leq 0^-$, respectively.
The surface impedance can be calculated as
\begin{equation}
 Z_i = -\frac{\jj \kappa^{(i)}}{\omega \varepsilon_y^{(i)}}.  \label{eq:83}
\end{equation}
The resonance condition $Z_1 + Z_2 = 0$ is reduced to
\begin{equation}
 -\frac{\kappa^{(1)}}{\varepsilon_y^{(1)}} =  \frac{\kappa^{(2)}}{\varepsilon_y^{(2)}}.  \label{eq:84}
\end{equation}
Because $\varepsilon_y^{(1)}>0$ and $\varepsilon_y^{(2)}<0$ are satisfied,
$\kappa^{(i)}$ can be positive.
From Eqs.~(\ref{eq:82}) and (\ref{eq:84}), we obtain
\begin{equation}
 k_0 = \frac{\omega}{c_0}=k_y \sqrt{
\frac{\epsilon_x^{(2)}\epsilon_y^{(2)} - \epsilon_x^{(1)}\epsilon_y^{(1)}}{\epsilon_x^{(1)}\epsilon_x^{(2)} \left(\epsilon_y^{(2)}-\epsilon_y^{(1)}\right) }}.  \label{eq:85}
\end{equation}
From Eqs.~(\ref{eq:82}) and (\ref{eq:85}), 
it is possible to confirm that
$\kappa^{(i)}> 0$, 
which indicates that the mode is localized at the boundary.

\section{Simulation Details on Temporal Boundary \label{sec:details_temporal_boundary}}

For Sec.~\ref{sec:temporal_boundary}, 
we explain the details of the simulation of the temporal boundary.
We focus on the sudden temporal shift from 
a single-metalized waveguide to a double-metalized waveguide,
as shown in Figs.~\ref{fig:temporal_boundary}(a) and (b).
The temporal transition of the boundary resistance
causes a dispersion change in the waveguide.
Owing to the spacetime continuity condition, 
the wave vector must be conserved at the temporal boundary. 
The incident field in the single-metalized waveguide is
redistributed to a series of different frequency modes 
in the double-metalized waveguide.
This process is decomposed into scattering from the incident wave to 
the $\ell$th mode.
The mode number of the double-metalized waveguide 
is denoted as $\ell=0,\ 1,\ 2,\ \cdots$, in order from the lowest frequency of the zero mode ($\ell=0$).
The mode with $\ell>0$ has the transverse wavenumber $k_x^{(\ell)}= \ell \pi/d$ in $x$.
The temporal dynamics were simulated using the transient analysis (temw) module in COMSOL Multiphysics.

\begin{figure}[!t]
 \includegraphics{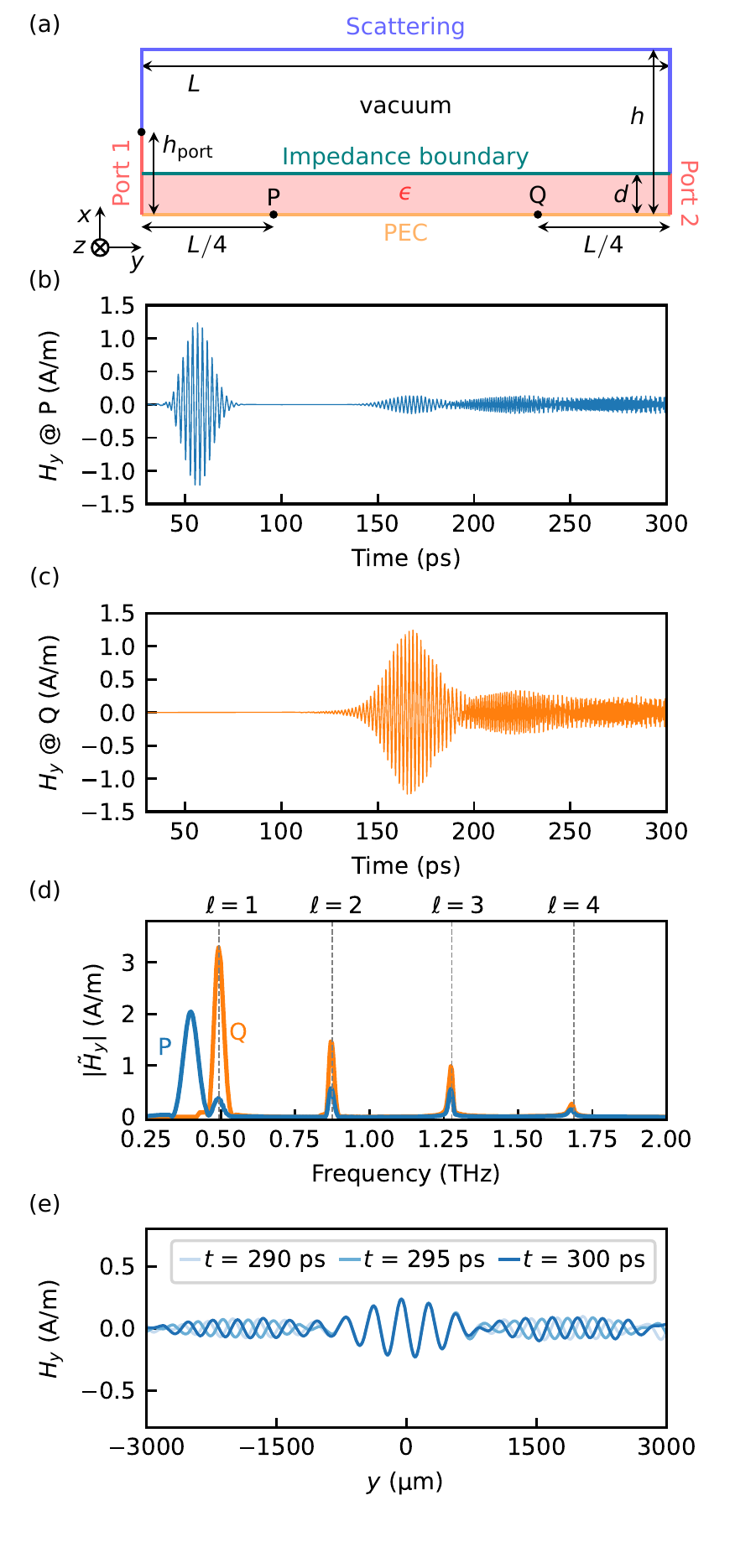}
  \caption{
Scattering at the temporal boundary:
 (a) Simulation setup. 
Temporal variation of $H_y$ at (c) P and (d) Q. (d) Amplitudes of Fourier-transformed signals at P and Q. The theoretical frequencies of the converted waves are depicted as
gray dashed lines.
(e) Spatial distribution of $H_y$ on $x=0$ for $t=290,\ 295,$ and $\SI{300}{ps}$.
\label{fig:temporal_boundary_detail}}
\end{figure}

The simulation setup is shown in Fig.~\ref{fig:temporal_boundary_detail}(a).
A dielectric slab made of GaAs with $\epsilon=12.96$ 
is placed in $(x,y) \in [0,d]\times [-L/2, L/2]$ under vacuum conditions.
The geometric parameters are set as $L=\SI{12}{mm}$, $h=\SI{3}{mm}$, 
and $d=\SI{100}{um}$.
The perfect electric conductor (PEC) boundary condition 
is imposed at the bottom of the slab ($x=0$).
The top side of the slab ($x=d$) is represented 
by the impedance boundary with a sheet admittance $Y_S$. 
$Y_S$ is initially set to 0 and is suddenly increased to 
$Y_S = \SI{1e10}{S}$ for $t\geq t_1 = \SI{100}{ps}$.
The wave packet is generated from Port 1 located in $\{(x,y)|0\leq x\leq h_\mathrm{port},\ y=-L/2 \}$ with $h_\mathrm{port}=4d$.
The incident sheet current $K_z$ along $z$ in Port 1 is 
given for $t\geq 0$ as follows:
\begin{multline}
 K_z/(2H_0) =\\
\begin{cases}
 \sin(\omega t) e^{-(t-t_0)^2/T^2}\sin(k_x x) & (0\leq x \leq d)\\
 \sin(\omega t) e^{-(t-t_0)^2/T^2}\sin(k_x d)e^{-\kappa (x-d)} & (x\geq d)
\end{cases}   \label{eq:86}
\end{multline}
The incident magnetic 
field $H_0$ is multiplied by the factor of 2 
considering that the incident current 
is equally divided between the output port resistance and the load, 
which are connected in parallel, under the matching condition.
We choose $H_0 = \SI{1}{A/m}$, $T=\SI{10}{ps}$, $t_0=\SI{13}{ps}$, and $\omega = 2\pi \times \SI{0.4}{THz}$, whereas $k_x$ and $\kappa$ are semi-analytically evaluated (see Supplemental Material in Ref.~\onlinecite{Miyamaru2021}).
The sheet impedance of the port inside the slab is 
set as that of the double-metalized waveguide, which suppresses the port reflection for the first-order mode.
Note that the boundary cannot absorb the other frequency modes.
For the outside region, we simply use the vacuum impedance $Z_0$.
In summary, the sheet impedance of the port is set as
\begin{equation}
 Z_p = 
\begin{cases}
 \frac{\mu_0 \omega^{(1)}}{k_y} & (0\leq x\leq d)\\
 Z_0 & (x\geq d)
\end{cases}   \label{eq:87}
\end{equation}
Although $Z_p$ is not correct for the single-metalized waveguide, 
it works practically. 
Here, $k_y$ is calculated from the dispersion of the single-metalized waveguide for the given $\omega$. The angular frequency $\omega^{(1)}$ of the $\ell=1$ mode in the double-metalized waveguide is evaluated for the calculated $k_y$.
Additionally, we place Port 2 at $\{(x,y)|0\leq x\leq d,\ y=L/2\}$.
No current is excited at Port 2, and 
its impedance is set as that of the double-metalized waveguide,
which is given by Eq.~(\ref{eq:87}). 
The remaining boundaries are treated as 
the scattering boundaries.
We monitor the magnetic fields at $(x,y)=(0, -L/4)$ and $(0, L/4)$, which are denoted as P and Q, respectively.

Figure~\ref{fig:temporal_boundary_detail}(b) shows the temporal variation of $H_y$ at P.
We can observe the incident wave packet at approximately $t=\SI{56}{ps}$ and the reflected signals after $t=\SI{170}{ps}$.
The first reflected signal at approximately $t=\SI{170}{ps}$ represents reflection into the $\ell=1$ mode.
By dividing the envelope amplitude of the first reflected $H_y$ by
that of the incident one,
the absolute value of the amplitude reflection coefficient is estimated 
as $0.11$, which agrees well with the theoretical value of $0.11$.
Because the higher-order modes propagate with lower group velocities,
they are observed later.
In fact, the oscillation frequency becomes significantly higher after $t=\SI{190}{ps}$.
The transmitted signal observed at Q is shown in Fig.~\ref{fig:temporal_boundary_detail}(c).
Similar behavior is observed for the reflected one.
The amplitude transmission coefficient of the $\ell=1$ mode is estimated as $1.0$, 
which agrees with the theoretical value of $1.0$.
To analyze the frequency shift, 
the signals in Figs.~\ref{fig:temporal_boundary_detail}(b) and (c) are 
Fourier-transformed with the normalized convention of unitary transformation.
The Fourier amplitude $|\tilde{H}_y|$ is shown in Fig.~\ref{fig:temporal_boundary_detail}(d). The peak frequencies agree well with the theoretically obtained frequencies of the converted waves, as indicated by vertical dashed gray lines.
To investigate the spatial distribution of the zero mode, 
we plotted $H_y$ along $x=0$ at $t=290,\ 295$, and $\SI{300}{ps}$, as shown in Fig.~\ref{fig:temporal_boundary_detail}(e).
The central signal at approximately $y=0$ is steady and represents the zero mode,
although the other region exhibits time evolution.
The envelope amplitude of the zero mode is divided by the incident one,
and we obtain the scattering coefficient of $0.19$ to the zero mode.
This value is identical to the theoretical value of $0.19$.
The results validate the simulation.

\section{Reciprocity in Electrostatics \label{sec:reciprocity}}
In this section, we summarize reciprocity in electrostatics, as discussed in Sec.~\ref{sec:uniform_layer}.
Consider two electrostatic potentials $\varphi_1(x,y,z)$ and $\varphi_2(x,y,z)$
for a reciprocal permittivity (represented by a scalar or a symmetric matrix).
The corresponding electric displacements are denoted as $\mathbf{D}_1$ and $\mathbf{D}_2$, respectively.
When we assume that there is no free charge,
the following reciprocal theorem for a three-dimensional region $\Omega$
holds:
\begin{multline}
 \int_{\partial \Omega} (\varphi_1 \mathbf{D}_2 - \varphi_2 \mathbf{D}_1 )\cdot \dd \mathbf{S}
= \int_{\Omega} \nabla\cdot (\varphi_1 \mathbf{D}_2 - \varphi_2 \mathbf{D}_1 ) \dd V\\
= \int_{\Omega}  (-\mathbf{E}_1 \cdot \mathbf{D}_2 + \mathbf{E}_2 \cdot \mathbf{D}_1) \dd V = 0.  \label{eq:88}
\end{multline}

We apply the reciprocity theorem for a slab presented in Sec.~\ref{sec:uniform_layer}.
Consider $\Omega = \left\{(x,y,z)|-d/2\leq x \leq d/2\right\}$
and assume $k_y \ne 0$.
For $k_y \ne 0$, the complex amplitude $\tilde{\varphi}$ of the electrostatic potential is proportional to $\tilde{E}_y$.
When the $\tilde{E}_y(\pm d/2)$ values are given, the electric displacements are linearly determined as follows:
\begin{equation}
 \begin{bmatrix}
  \tilde{D}_x\left(\frac{d}{2}\right)\\
  -\tilde{D}_x\left(-\frac{d}{2}\right)
 \end{bmatrix}
=
\begin{bmatrix}
 \alpha_{11} & \alpha_{12}\\
 \alpha_{21} & \alpha_{22}
\end{bmatrix}
 \begin{bmatrix}
  \tilde{E}_y\left(\frac{d}{2}\right)\\
  \tilde{E}_y\left(-\frac{d}{2}\right)
 \end{bmatrix}   \label{eq:89}
\end{equation}
Let $\tilde{\varphi}_1$ be the complex amplitude of the electrostatic potential for $\tilde{E}_y(d/2) = 1$ and $\tilde{E}_y(-d/2) = 0$.
Similarly, $\tilde{\varphi}_2$ is defined for
$\tilde{E}_y(d/2) = 0$ and $\tilde{E}_y(-d/2) = 1$.
By applying Eq.~(\ref{eq:88}) for $\tilde{\varphi}_1$ and $\tilde{\varphi}_2$, we obtain $\alpha_{12}=\alpha_{21}$.
Finally, $\det F_0 =1$ is obtained, 
by transforming Eq.~(\ref{eq:89}) into the form of Eq.~(\ref{eq:34}).

\section{Formal Solution for Layered Medium \label{sec:formal}}

For Sec.~\ref{sec:nonuniform}, we 
construct a formal solution of an electrostatic field.
Consider $\varepsilon(x)$ in $x\geq 0$.
From Eqs.~(\ref{eq:60}) and (\ref{eq:63}) with zero frequency,
the following equation determines the spatial variation of the field:
\begin{equation}
-\jj \odv{}{x}
\begin{bmatrix}
 \tilde{D}_x\\
 \tilde{E}_y
\end{bmatrix}
=  k_y
\begin{bmatrix}
 0 & \varepsilon(x)\\
 -\varepsilon(x)^{-1} & 0
\end{bmatrix}
\begin{bmatrix}
 \tilde{D}_x\\
 \tilde{E}_y
\end{bmatrix}.   \label{eq:90}
\end{equation}
Equation~(\ref{eq:90}) is analogous to 
Schr\"{o}dinger's equation with a time-dependent Hamiltonian.
When we assume $\varepsilon_0(x)=\varepsilon_0$ for $x\geq x_0$, 
the solution may satisfy
\begin{equation}
 \begin{bmatrix}
  \tilde{D}_x(x_0)\\
  \tilde{E}_y(x_0)
 \end{bmatrix}
=
 \begin{bmatrix}
  \varepsilon_0\\
  \jj
 \end{bmatrix}.  \label{eq:91}
\end{equation}
Let $A(x)$ be a 2$\times 2$ matrix defined at $x \in [0,\infty)$.
The spatially ordering operator $\mathbb{S}$ acts as follows:
\begin{equation}
\mathbb{S} \left[A(x_1)A(x_2) \cdots A(x_n)\right]
:= 
A(x_{\sigma(1)})A(x_{\sigma(2)}) \cdots A(x_{\sigma(n)})  \label{eq:92}
\end{equation}
with a permutation $\sigma \in S_n$ that satisfies
$x_{\sigma(1)} \geq x_{\sigma(2)} \geq \cdots \geq x_{\sigma(n)}$.
Using $\mathbb{S}$, 
we can formally express the field at any point $x$ ($0\leq x \leq x_0$) as follows:
\begin{multline}
 \begin{bmatrix}
 \tilde{D}_x (x)\\
 \tilde{E}_y (x)
\end{bmatrix} \\
=
\left[\mathbb{S}\exp\left(\jj k_y \int^{x_0}_x \dd x'\begin{bmatrix}
 0 & \varepsilon(x')\\
 -\varepsilon(x')^{-1} & 0
\end{bmatrix}
\right)\right]^{-1}
\begin{bmatrix}
 \varepsilon_0\\
 \jj
\end{bmatrix}.   \label{eq:93}
\end{multline}

\section{Topological Polarization Rotation in Periodic Layered Media\label{sec:prf_tp_rot_per}}

For Sec.~\ref{sec:nonuniform}, we provide proof of topological polarization 
rotation in periodic media with infinite numbers of layers.
We select $x_0 = 0 < x_1 < x_2 < \cdots <x_n $ and place a dielectric material with $\varepsilon_i>0$ (relative permittivity: $\epsilon_i$) in $x\in [x_{i-1}, x_i]$.
We regard $[x_0, x_n]$ as a unit cell
and arrange it periodically in $x\geq 0$ with period $a=x_n-x_0$.
Let us consider a TM wave with wavenumber $k_y>0$ in $y$.
First, we show that $\tilde{E}_y(0)/\tilde{D}_x(0)$ 
is purely imaginary at $x=0$.
The F$_0$ matrix $F^\mathrm{(unit)}_0$ of the unit cell has the following form: 
\begin{equation}
\begin{bmatrix}
 \alpha & - \jj \beta\\
 \jj \gamma & \delta
\end{bmatrix}, \
\alpha>1,\ \beta>0,\ \gamma>0,\ \delta>1,\  
 \alpha \delta -\beta \gamma =1.   \label{eq:94}
\end{equation}
The above statement is proved as follows:
For $\varepsilon>0$, Eq.~(\ref{eq:35}) fulfills Eq.~(\ref{eq:94}).
If we multiply Eq.~(\ref{eq:94}) by the matrix in Eq.~(\ref{eq:35}),
the obtained matrix also satisfies Eq.~(\ref{eq:94}).
Therefore, the statement is proven.
From Eq.~(\ref{eq:94}),
the eigenvalue $\Lambda$ of $F_0^\mathrm{(unit)}$ is calculated as follows:
\begin{equation}
 \Lambda = \frac{(\alpha+\delta)\pm \sqrt{(\alpha+\delta)^2-4} }{2}.   \label{eq:95}
\end{equation}
Therefore, $\Lambda$ must be real.
Furthermore, $\tilde{E}_y(0)/\tilde{D}_x(0)$ becomes purely imaginary as follows:
\begin{equation}
\frac{\tilde{E}_y(0)}{\tilde{D}_x(0)} = \frac{\tilde{E}_y(a)}{\tilde{D}_x(a)} = \frac{\alpha-\Lambda}{\jj\beta}.  \label{eq:96}
\end{equation}

Now, we consider an electrostatic field in $x\geq 0^+$
and assume that $[\tilde{D}_x(0)\ \tilde{E}_y(0)]^\mathrm{T} = [\varepsilon_C\ - \jj\xi]^\mathrm{T}$ ($\varepsilon_C > 0$, $\xi \geq 0$).
We define $d_1=x_1 -x_0$ and calculate the field at $x=x_1$ as follows:
\begin{equation}
 \begin{bmatrix}
  \tilde{D}_x(x_{1})\\ \tilde{E}_y(x_{1})
 \end{bmatrix}
=
\begin{bmatrix}
 \cosh(k_y d_1) \varepsilon_C + \varepsilon_1 \xi\sinh (k_yd_1)\\
-\jj\left(\frac{\varepsilon_C}{\varepsilon_1} \sinh(k_yd_1) + \xi \cosh(k_yd_1)\right)
\end{bmatrix}.   \label{eq:97}
\end{equation}
This field has the form of $[\varepsilon_C\ - \jj\xi]^\mathrm{T}$ ($\varepsilon_C> 0$, $\xi\geq 0)$. 
Moreover, Eq.~(\ref{eq:97}) indicates that $\tilde{D}_x(x_1) > \tilde{D}_x(0)$.
Repeating the discussion, we can express the field in 
the form of $[\varepsilon_C\ - \jj\xi]^\mathrm{T}$ ($\varepsilon_C>0$, $\xi\geq 0$)
at \textit{any} point. 
Moreover, $\tilde{D}(x)$ does not converge at $x\rightarrow +\infty$.
Therefore, it does not give a solution.
In conclusion, the field can be expressed in the form of 
$[\varepsilon_C\ \jj\xi]^\mathrm{T}$ ($\varepsilon_C> 0$, $\xi> 0)$
at \textit{any} point of $x\geq 0$, where we used a degree of freedom of the overall scalar factor.

\section{Analysis of Ladder Circuits \label{sec:ladder_analysis}}

\begin{figure}[t]
 \includegraphics{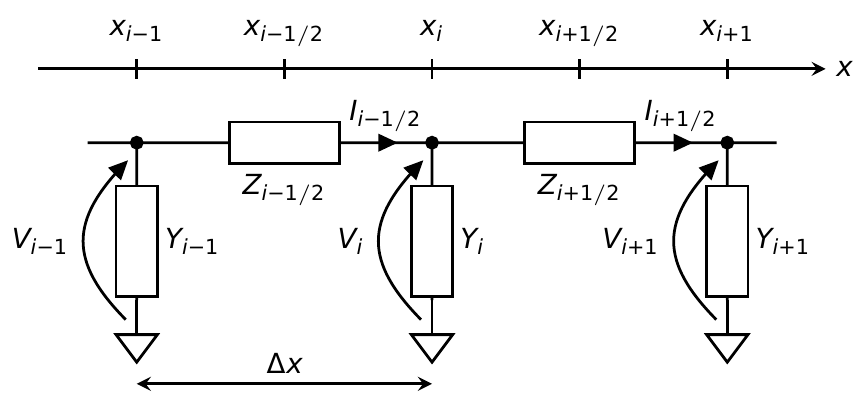}
  \caption{
Transmission line with admittance $Y_i$ and impedance $Z_{i+1/2}$
 along the $x$-axis.
\label{fig:transmission_line}}
\end{figure}

For Sec.~\ref{sec:dielectric_metal}, 
we consider a one-dimensional circuit, as shown in Fig.~\ref{fig:transmission_line}.
The $x$-axis is discretized with $\Delta x$, and the
shunt admittance $Y_i$ and series impedance $Z_{i+1/2}$
are located at $x_i = i \Delta x$ and $x_{i+1/2} = (i+1/2)\Delta x$ ($i\in \mathbb{Z}$), respectively. 
The complex amplitudes of the current $\tilde{I}_{i+1/2}$ flowing through $Z_{i+1/2}$
and those of the voltage $\tilde{V}_i$ along $Y_i$ satisfy the following equations:
\begin{align}
\tilde{I} _{i+\frac{1}{2}} -\tilde{I} _{i-\frac{1}{2}} &= - Y_i \tilde{V}_i,  \label{eq:98}\\
\tilde{V} _{i+1} -\tilde{V} _{i} &= - Z_{i+\frac{1}{2}} \tilde{I}_{i+\frac{1}{2}}.  \label{eq:99}
\end{align}

For a periodic system, 
we assume that $Y_i$ and $Z_{i+1/2}$ are independent of $i$ and denote them as $Y$ and $Z$, respectively.
With the Bloch wavenumber $k_x^\mathrm{(Bloch)}$, 
Bloch's theorem indicates that
\begin{align}
 \tilde{V}_{i+1} &= \exp(-\jj k_x^\mathrm{(Bloch)} \Delta x) \tilde{V}_{i},  \label{eq:100}\\
 \tilde{I}_{i-\frac{1}{2}} &= \exp(\jj k_x^\mathrm{(Bloch)} \Delta x) \tilde{I}_{i+\frac{1}{2}}.  \label{eq:101}
\end{align}
By substituting Eqs.~(\ref{eq:100}) and (\ref{eq:101}) into Eqs.~(\ref{eq:98}) and (\ref{eq:99}),
we obtain
\begin{equation}
 4\sin^2\left(\frac{k_x^\mathrm{(Bloch)} \Delta x}{2}\right) = -ZY.  \label{eq:102}
\end{equation}
For $k_x^\mathrm{(Bloch)}=0$, Eq.~(\ref{eq:102}) is reduced to $YZ=0$, 
which indicates the existence of symmetric ($Y=0$) and antisymmetric ($Z=0$) modes.
Equation~(\ref{eq:102}) is often used to calculate a dispersion relation $\omega(k_x^\mathrm{(Bloch)})$ for a given $k_x^\mathrm{(Bloch)}$.
Instead of Eq.~(\ref{eq:102}), we can also determine $\exp(-jk_x^\mathrm{(Bloch)}\Delta x)$ as
\begin{equation}
\exp(-jk_x^\mathrm{(Bloch)}\Delta x) = \frac{YZ+2 \pm \sqrt{YZ(YZ+4)}}{2}.  \label{eq:103}
\end{equation}
Equation~(\ref{eq:103}) is utilized to obtain a dispersion relation $k_x^\mathrm{(Bloch)}(\omega)$ for a given $\omega$.

\begin{figure}[t]
 \includegraphics{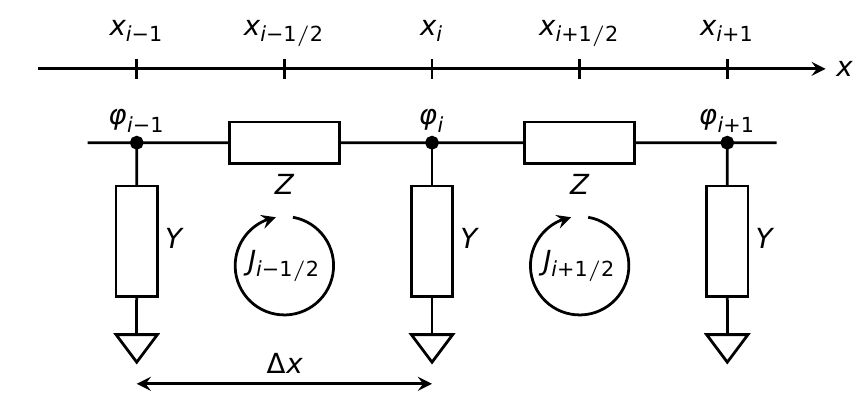}
  \caption{Node potential $\varphi_i$ and mesh current $J_{i+1/2}$ in a uniform transmission line.
\label{fig:constant_transmission_line}}
\end{figure}

We can intuitively understand the eigenmode symmetry
at the highly symmetric points in the Brillouin zone.
As shown in Fig.~\ref{fig:constant_transmission_line}, we define node potentials $\{\varphi_i\}$ and mesh currents $\{J_{i+1/2}\}$, which represent the full degree of freedom restricted by Kirchhoff's laws \cite{Bamberg1990,Nakata2012a,Nakata2019a}.
Additionally, we introduce two mirror reflections $\mathcal{M}_{x_i}$
and $\mathcal{M}_{x_{i+1/2}}$ on $x=x_i$ and $x_{i+1/2}$, respectively.
First, we analyze the $k_x^\mathrm{(Bloch)}=0$ case,
in which $\varphi_i$ and $J_{i+1/2}$ are independent of $i$.
Clearly, the constant node potential represents 
a symmetric degree of freedom (for both $\mathcal{M}_{x_i}$ and $\mathcal{M}_{x_{i+1/2}}$), 
whereas the constant mesh current represents an antisymmetric one.
Therefore, symmetric and antisymmetric modes can appear at $k_x^\mathrm{(Bloch)}=0$.
Next, we examine the $k_x^\mathrm{(Bloch)}\Delta x=\pm\pi$ case.
Here, $(-1)^i \varphi_i$ and $(-1)^i J_{i+1/2}$ are constants regardless of $i$.
Interestingly, both degrees of freedom produce 
the $\mathcal{M}_{x_i}$-symmetric 
($\mathcal{M}_{x_{i+1/2}}$-antisymmetric) distributions.
Therefore, the modes at $k_x^\mathrm{(Bloch)}\Delta x=\pm\pi$ have identical parity.

\begin{figure}[t]
 \includegraphics{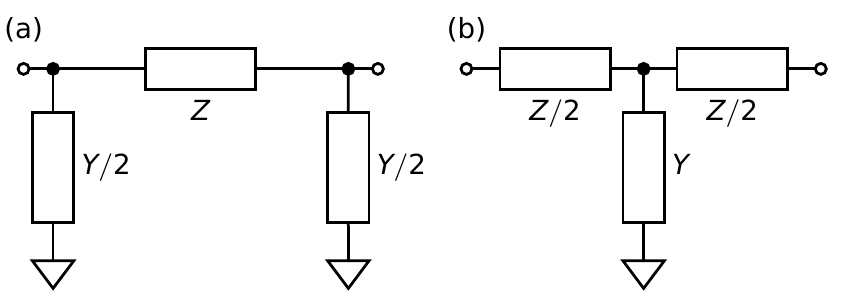}
  \caption{
\label{fig:mirror_symmetric_circuit_unit} (a) $\Pi$ and (b) T units. }
\end{figure}

The Bloch impedance is the effective impedance of infinite circuits
seen from left to right.
It depends on the choice of a unit cell.
Here, we consider two configurations of unit cells, as shown in Fig.~\ref{fig:mirror_symmetric_circuit_unit}.
For the $\Pi$ unit, we can evaluate the Bloch impedance as
\begin{align}
 Z^\mathrm{(Bloch)}_\Pi &= \frac{2Z}{YZ + 2[1 - \exp(-\jj k_x^\mathrm{(Bloch)}\Delta x)]}  \nonumber\\
&=\mp \frac{2Z}{\sqrt{YZ(YZ+4)}}.  \label{eq:104}
\end{align}
It is also calculated for the T unit as
\begin{align}
 Z^\mathrm{(Bloch)}_T  &= \frac{Z}{2} + \frac{1-\exp(-\jj k_x^\mathrm{(Bloch)}\Delta x)}{Y} \nonumber\\
&=\mp\frac{\sqrt{YZ(YZ+4)}}{2Y}.  \label{eq:105}
\end{align}
The circuit duality connects these impedances as follows:
\begin{equation}
 Z^\mathrm{(Bloch)}_\Pi Z^\mathrm{(Bloch)}_T = \frac{Z}{Y}.  \label{eq:106}
\end{equation}
Note that 
$\operatorname{Im}Z^\mathrm{(Bloch)}_\Pi$
and 
$\operatorname{Im}Z^\mathrm{(Bloch)}_T$
can have different signs
when we assume that $Z/Y$ is real.

Finally, we consider a CRLH transmission line,
as shown in Fig.~\ref{fig:CRLH}(a).
The series impedance $Z$ is expressed as
\begin{equation}
 Z = \jj \omega L_{R} + \frac{1}{\jj \omega C_{L}},  \label{eq:107}
\end{equation}
and the shunt admittance is given as
\begin{equation}
 Y = \jj \omega C_{R} + \frac{1}{\jj \omega L_{L}}.  \label{eq:108}
\end{equation}
It is not essential to distinguish which capacitor or inductor
is on the left side in $Z$, considering that the effective response 
is unaffected by the interchange of the geometrical positions.
We define parameters $\zeta:=C_{L}/C_{R}$, 
$\eta = L_{L}/L_{R}$, $\omega_0=1/\sqrt{L_{R}C_{R}}$, and
$R_0 =\sqrt{L_{R}/C_{R}}$.
Then, $Z$ and $Y$ are expressed as
\begin{align}
 Z(\omega/\omega_0, \zeta, R_0)&=R_0\left(\jj \frac{\omega}{\omega_0} + \frac{1}{j\zeta \frac{\omega}{\omega_0}}\right),  \label{eq:109}\\
Y(\omega/\omega_0,\eta, R_0)&={R_0}^{-1}\left(\jj \frac{\omega}{\omega_0} + \frac{1}{j\eta \frac{\omega}{\omega_0}} \right).  \label{eq:110}
\end{align}
From Eq.~(\ref{eq:102}), the dispersion relation can be calculated as follows:
\begin{equation}
 \left[2\sin\left(\frac{k_x^\mathrm{(Bloch)} \Delta x}{2}\right)\right]^2 = \left(\frac{\omega}{\omega_0}\right)^2
+ \frac{1}{\zeta \eta}\left(\frac{\omega_0}{\omega}\right)^2 - \frac{1}{\zeta}- \frac{1}{\eta}.  \label{eq:111}
\end{equation}
Using Eq.~(\ref{eq:103}), we can calculate the complex dispersion relation.

\section{Symmetry Constraints \label{sec:sym_constraint}}

For Sec.~\ref{sec:dielectric_metal}, 
we summarize the symmetry constraints on the wavenumber and Bloch impedance.
Consider TM wave propagation in a layered photonic or plasmonic
crystal periodic in $x$ with period $a$.
One example is the model examined in Appendix~\ref{sec:periodic}.
Here, we derive the symmetry constraints on the Bloch wavenumber and impedance.
The same constraints also work in circuits.
The assumption of the TM mode is 
introduced to simplify the explanation and is not essential.

\paragraph{Time-Reversal Symmetry and Reciprocity.}

First, we discuss the time-reversal operation.
For an electric field $E(t) = \tilde{E}\exp(\jj \omega t) + \mathrm{c.c.}$, $\tilde{E}^*$ gives the time-reversal phasor of $E(-t)$.
The phasor $\tilde{H}$ of a magnetic field is transformed into $-\tilde{H}^*$.
Therefore, the time-reversal operation for $[\tilde{E}\ \tilde{H}]^\mathrm{T}$
is represented by $\mathcal{J}= K \sigma_z$, where $K$ is the complex-conjugate operator and $\sigma_z = \mathrm{diag}(1,-1)$.
If the system has time-reversal symmetry, $\tilde{E}^*$ and $-\tilde{H}^*$ give the solution to the problem.

\begin{figure}[bt]
 \includegraphics{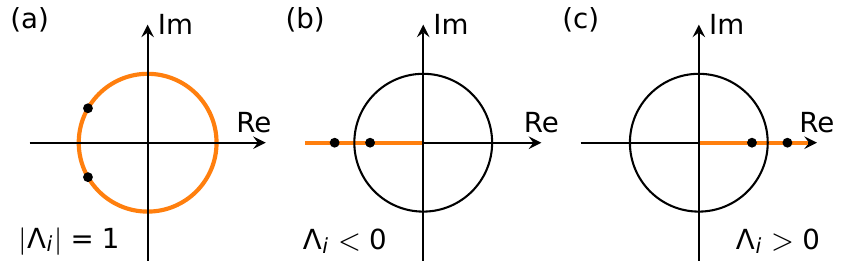}
  \caption{
\label{fig:F_eigenvalue} Time-reversal symmetry and reciprocity constraint
locations of eigenvalues $\Lambda_i = \exp(-\jj k_x^\mathrm{(Bloch)} a)$ in the complex plane. 
The eigenvalue locations are classified into (a) propagating and (b) and (c) decaying (in-gap) modes. Here, $a$ represents the period in $x$.}
\end{figure}

A real $\varepsilon$ and $\mu$ make $F$ in Eq.~(\ref{eq:69})
invariant under the time-reversal operation.
Let us examine Eq.~(\ref{eq:74}).
The two eigenvalues of $F^\mathrm{(unit)}$ are denoted as $\Lambda_1$ and $\Lambda_2$.
From $\det F^\mathrm{(unit)}=1$, according to the reciprocity, 
$\Lambda_1$ and $\Lambda_2$ depend on each other as
$\Lambda_1\Lambda_2 =1$.
For $\Lambda_1 \ne \Lambda_2$, the time-reversal symmetry demands (i) $\Lambda_1 = \Lambda_1^*$, $\Lambda_2 = \Lambda_2^*$ or (ii) $\Lambda_1 = \Lambda_2^*$.
From these constraints,
the distribution of $\Lambda$ is classified in Fig.~\ref{fig:F_eigenvalue},
where $\pm 1$ gives the crossover points between propagating and decaying (in-gap) solutions.
The time-reversal operation does not change the eigenvalue $\Lambda$
of the decaying solutions. 
Therefore, we obtain $Z^\mathrm{(Bloch)} = -(Z^\mathrm{(Bloch)})^*$, which indicates that the Bloch impedance inside a band gap is purely imaginary.

\paragraph{Mirror Symmetry.}

Next, we consider a mirror-symmetric unit cell with
time-reversal symmetry and reciprocity.
Figure~\ref{fig:dielectric_metal_transition}(a) shows an example.
On the mirror-symmetric plane, there are constraints on the Bloch impedance.

First, consider a propagating mode with angular frequency $\omega>0$ 
and a real $k_x^\mathrm{(Bloch)}$.
The invariance under a combination of the mirror reflection and time-reversal operations
leads to $Z^{\mathrm{(Bloch)}} = (Z^{\mathrm{(Bloch)}})^*$. 
Note that the mirror reflection induces the transformation
$\tilde{H}_z\rightarrow -\tilde{H}_z$,
whereas the time-reversal operation induces $\tilde{H}_z\rightarrow -\tilde{H}^*_z$.
Therefore, the Bloch impedance is real in propagating bands.

Second, we show that $Z^\mathrm{(Bloch)} = \infty$ or $0$ on a mirror plane
for $k_x^\mathrm{(Bloch)} = 0$ and $\pm \pi/a$. Here, the period in $x$ is denoted as $a$.
Because $k_x^\mathrm{(Bloch)} = 0$ and $\pm \pi/a$ are invariant under the mirror reflection,
the eigenmodes must be symmetric or antisymmetric with respect to the mirror reflection.
Owing to the field continuity, 
the symmetric and antisymmetric solutions must satisfy $\tilde{H}_z=0$ 
and $\tilde{E}_y=0$ on the mirror plane, 
which results in $Z^\mathrm{(Bloch)} = \infty$ and $0$, respectively \cite{Xiao2014}.

Third, the reverse of the second statement holds: 
$Z^\mathrm{(Bloch)} = \infty$ and $Z^\mathrm{(Bloch)} = 0$ 
at a propagating band 
indicate symmetric and antisymmetric modes, respectively.
We prove this statement for $Z^\mathrm{(Bloch)} = \infty$.
The $x$-axis is selected such that the unit cell $\left\{(x,y,z)|x\in [-a/2,a/2]\right\}$ has the mirror planes located on $x=0$ and $x=\pm a/2$.
The eigenvector of the F matrix can be selected as $[\tilde{E}\ \tilde{H}]^\mathrm{T}=[1\ 0]$,
which has time-reversal symmetry.
Therefore, the wavenumber is restricted to a time-reversal wavenumber
of $k_x^\mathrm{(Bloch)} = 0,\ \pm \pi/a$; i.e.,\
the field is symmetric or antisymmetric on the unit boundary $x=\pm a/2$.
Therefore, all fields inside the unit cell must be symmetric or antisymmetric.
A similar discussion holds for $Z^\mathrm{(Bloch)} = 0$.

Finally, we establish a definite Bloch-reactance sign in a band gap.
The band gap is denoted as $\omega_1<\omega< \omega_2$.
The Bloch impedance must be real in a propagating band,
whereas it is purely imaginary in a band gap. 
Therefore, $Z^\mathrm{(Bloch)}(\omega_i)$ must be $0$ or $\infty$ for $i=1,\ 2$, which indicates the existence of symmetric or antisymmetric modes.
We can safely assume that there is no zero or pole of $Z^\mathrm{(Bloch)}$ in $\omega_1<\omega<\omega_2$.
Even if there was a zero or pole in $\omega_1<\omega<\omega_2$,
we would divide the band gap into several segments, each of which fulfills the assumption.
Therefore, the reactance theorem maintain the definite reactance sign in each band gap.

\section{Analogy Between Circuits and Electromagnetic Systems \label{sec:analogy_circuit_EM}}

For Sec.~\ref{sec:CRLH_dielectric_metal}, we present an analogy between electric circuits and electromagnetic systems.
We consider TM-wave propagation along $x$ with $k_y=0$.
The electric field is parallel to the $y$-axis,
and the magnetic field is in the $z$ direction.
These complex amplitudes are denoted as $\tilde{E}_y(x)$ and $\tilde{H}_z(x)$.
The electric permittivity and magnetic permeability are represented by
$\varepsilon(x)$ and $\mu(x)$, respectively.
Amp\`{e}re--Maxwell's and Faraday's laws give the following equations with angular frequency $\omega$:
\begin{align}
 \odv{\tilde{H}_z}{x} &= -\jj\omega \varepsilon \tilde{E}_y,  \label{eq:112}\\
 \odv{\tilde{E}_y}{x} &= -\jj\omega \mu \tilde{H}_z.  \label{eq:113}
\end{align}
Clearly, Eqs.~(\ref{eq:112}) and (\ref{eq:113}) are analogous to Eqs.~(\ref{eq:98})
and (\ref{eq:99}), respectively. 
The variable correspondence between the two systems is summarized in Tab.~\ref{tab:analogy}.

\begin{table}[htbp]
\renewcommand{\arraystretch}{1.5}
  \caption{\label{tab:analogy} Analogy between circuits and electromagnetic systems.}
  \centering
  \begin{tabular}{c||c|c|c|c}
Circuit& $\tilde{V}_{i}$ & $\tilde{I}_{i+1/2}$& $Y_i$ & $Z_{i+1/2}$\\ \hline
Maxwell&$\tilde{E}_y\Delta x$ &$\tilde{H}_z \Delta x$ & $\jj\omega \varepsilon \Delta x$&
$\jj\omega \mu \Delta x$
  \end{tabular}
\end{table}

\section{Circuit Models of Dielectric and Metal \label{sec:dielectric_metal_models}}

For Sec.~\ref{sec:CRLH_dielectric_metal}, 
we construct circuit models of the dielectric and metal
to reproduce their wavenumber ($k_x$) and wave impedance ($Z$).

\paragraph{Effective Parameters for TM Propagation.}
For the TM wave with $k_y$ considered in Appendix~\ref{sec:basicEqs},
we can construct a one-dimensional model along $x$.
The effective permittivity and permeability in the one-dimensional model are 
denoted as $\varepsilon_\mathrm{eff}$ and $\mu_\mathrm{eff}$, respectively.
From Eq.~(\ref{eq:67}),
we obtain
\begin{equation}
 \varepsilon_\mathrm{eff} = \frac{k_x}{\omega Z}.  \label{eq:114}
\end{equation}
Considering $k_y=0$ in Eq.~(\ref{eq:65}),
$k_x$ should satisfy 
\begin{equation}
{k_x}^2 = \varepsilon_\mathrm{eff} \mu_\mathrm{eff} \omega^2.  \label{eq:115}
\end{equation}
Thus, $\mu_\mathrm{eff}$ can be obtained as
\begin{equation}
 \mu_\mathrm{eff} = \frac{k_x Z}{\omega}.  \label{eq:116}
\end{equation}

\paragraph{Circuit Model of Dielectric.}

\begin{figure}[t]
 \includegraphics{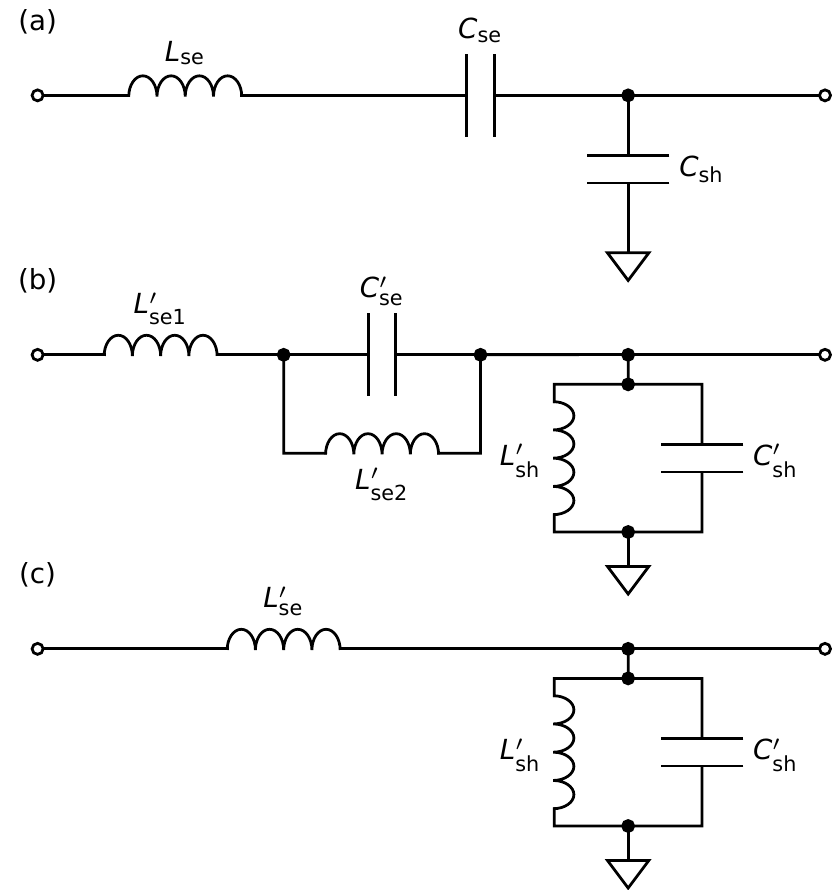}
  \caption{
\label{fig:transmission_line_models} Circuit models of (a) a dielectric and (b) a metal for transverse-magnetic waves. Under $\omega\ll \omega_p$, (b) can be approximated as (c).}
\end{figure}

We characterize a dielectric slab 
with permittivity $\varepsilon>0$ and vacuum permeability $\mu_0$.
Eqs.~(\ref{eq:67}) and (\ref{eq:71})
give the impedance and wavenumber, respectively.
Using Eq.~(\ref{eq:114}) and (\ref{eq:116}), we can extract the effective parameters as follows:
\begin{align}
 \varepsilon_\mathrm{eff} &= \varepsilon,  \label{eq:117}\\
 \mu_\mathrm{eff} &= \mu_0 \left[1-\left(\frac{c k_y}{\omega}\right)^2\right],  \label{eq:118}
\end{align}
where we define the speed of light in the slab as $c= 1/\sqrt{\varepsilon \mu_0}$.
Thus, the slab has a magnetic response, 
which is represented by the Drude frequency dispersion.
Therefore, the slab has a cutoff angular frequency of $c k_y$.
This configuration is the magnetic dual to
parallel metallic waveguides with transverse-electric (TE) modes.

Using the correspondence developed in Appendix~\ref{sec:analogy_circuit_EM}, 
we can construct a circuit model, as shown in Fig.~\ref{fig:transmission_line_models}(a).
The circuit parameters are determined as
\begin{align}
 L_\mathrm{se} &= \mu_0 \Delta x,  \label{eq:119}\\
 C_\mathrm{se} &= \frac{1}{(ck_y)^2 \mu_0 \Delta x},  \label{eq:120}\\
 C_\mathrm{sh} &= \varepsilon \Delta x.  \label{eq:121}
\end{align}
Here, $Z$ comprises $L_\mathrm{se}$ and $C_\mathrm{se}$, 
which induces the Drude-like response in Eq.~({\ref{eq:118}})
as a series resonance.

\paragraph{Circuit Model of Metal.}

Next, consider a Drude metal with 
the TM impedance $Z_m = -\jj \sqrt{{k_y}^2 -\varepsilon_m\mu_0 \omega^2}/(\omega \varepsilon_m)$ obtained from Eq.~(\ref{eq:18}).
The one-dimensional circuit model for the Drude metal is shown in Fig.~\ref{fig:transmission_line_models}(b).
The correspondence presented in Appendix~\ref{sec:analogy_circuit_EM} gives the following circuit parameters:
\begin{align}
 L_{\mathrm{se}1}' &=  \mu_0 \Delta x,  \label{eq:122}\\
 L_{\mathrm{se}2}' &= \frac{{k_y}^2}{{\omega_p}^2\varepsilon_0}\Delta x,  \label{eq:123}\\
 C_\mathrm{se}' &= \frac{\varepsilon_0}{{k_y}^2 \Delta x},  \label{eq:124}\\
 C_\mathrm{sh}' &= \varepsilon_0\Delta x,  \label{eq:125}\\
 L_\mathrm{sh}' &= \frac{1}{{\omega_p}^2 \varepsilon_0 \Delta x}.  \label{eq:126}
\end{align}

The impedance ratio between $Z'_{C\mathrm{se}} = 1/(\jj\omega C_\mathrm{se}')$ and $Z'_{L\mathrm{se}2} =\jj \omega L_{\mathrm{se}2}'$ is evaluated as
\begin{equation}
\left|\frac{Z'_{L\mathrm{se}2}}{Z'_{C\mathrm{se}}}\right| = \left(\frac{\omega}{\omega_p}\right)^2.  \label{eq:127}
\end{equation}
Therefore, $C_\mathrm{se}'$ can be regarded as open if $\omega\ll \omega_p$ is satisfied.
In this case, the model is reduced to Fig.~\ref{fig:transmission_line_models}(c),
where $L_\mathrm{se}'$ is given as
\begin{equation}
 L_\mathrm{se}' = L'_{\mathrm{se}1} + L'_{\mathrm{se}2}.  \label{eq:128}
\end{equation}

\section{Swapping Between Series and Shunt Elements \label{sec:YZswap}}

\begin{figure}[!b]
 \includegraphics{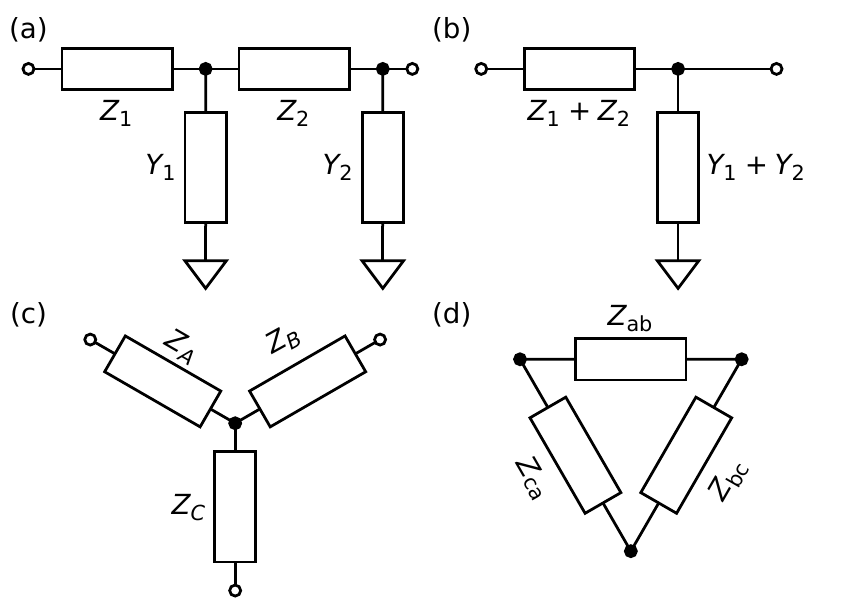}
  \caption{ (a) Unit cell; (b) approximated unit cell; (c) Y circuit; (d) $\Delta$ circuit.
\label{fig:YZ_swap}}
\end{figure}

For Sec.~\ref{sec:CRLH_dielectric_metal}, we summarize the swapping conditions between series and shunt elements.
Consider a unit cell, as shown in Fig.~\ref{fig:YZ_swap}(a).
When the shunt current is small, we can swap the positions of series and shunt elements
and obtain a unit cell, as shown in Fig.~\ref{fig:YZ_swap}(b).
To justify this transformation, we use
the Y--$\Delta$ transformation, which converts the Y circuit [Fig.~\ref{fig:YZ_swap}(c)] 
into the $\Delta$ circuit [Fig.~\ref{fig:YZ_swap}(d)].
The circuit parameters are related through the following equations:
\begin{align}
 Z_\mathrm{ab} &= \frac{Z_A Z_B + Z_B Z_C+Z_C Z_A}{Z_C},  \label{eq:129}\\
 Z_\mathrm{bc} &= \frac{Z_A Z_B + Z_B Z_C+Z_C Z_A}{Z_A},  \label{eq:130}\\
 Z_\mathrm{ca} &= \frac{Z_A Z_B + Z_B Z_C+Z_C Z_A}{Z_B}.  \label{eq:131}
\end{align}
Now, set $Z_\mathrm{A}=Z_1$, $Z_\mathrm{B}=Z_2$, and $Z_C = {Y_1}^{-1}$. 
Assuming $|Y_1|\ll |{Z_1}^{-1} + {Z_2}^{-1}|$,
we obtain $Z_\mathrm{ab} \approx Z_1 +Z_2$ and ${Z_\mathrm{ca}}^{-1} + {Z_\mathrm{bc}}^{-1}=Y_1$.
Then, the position of the unit cell is replaced, 
and we can justify the approximation.
Considering the dual circuit,
$|Z_i|\ll |{Y_1}^{-1} + {Y_2}^{-1}|$
gives another condition.

%

\end{document}